\documentclass[structabstract]{aa1}

\usepackage{savesym}
\usepackage{txfonts}
\savesymbol{iint}
\savesymbol{iiint}
\savesymbol{iiiint}
\savesymbol{idotsint}
\usepackage{natbib}
\bibpunct{(}{)}{;}{a}{}{,}
\usepackage{graphicx}
\usepackage{amsmath}
\usepackage{amssymb}
\usepackage{subfig}

\newcommand\units[1]{\nobreak\mbox{$\;$#1}}
\newcommand\percmcub{\nobreak\mbox{$\;$cm$^{-3}$}}
\newcommand\percmsq{\nobreak\mbox{$\;$cm$^{-2}$}}
\newcommand\fluxunit{\nobreak\mbox{$\;$erg$\;$cm$^{-2}\;$s$^{-1}$}}
\newcommand\fluxspunit{\nobreak\mbox{$\;$erg$^{-1}\;$cm$^{-2}\;$s$^{-1}$}}
\newcommand\heatunit{\nobreak\mbox{$\;$erg$\;$cm$^{-3}\;$s$^{-1}$}}

\newcommand\tento[1]{\nobreak\times\nobreak10^{#1}}

\def\apjl{Astrophysical Journal, Letters}
\def\apjs{Astrophysical Journal, Supplement}
\def\aap{Astronomy and Astrophysics}

\def\newblock{\hskip .11em \@plus .33em \@minus .07em}

\begin{document}

\title{Stationary and impulsive injection\\ of electron beams in converging magnetic field}

\titlerunning{Stationary and impulsive injection of electron beams}

\author{Taras V. Siversky \and Valentina V. Zharkova}

\authorrunning{T. V. Siversky \& V. V. Zharkova}

\institute{Department of Mathematics, University of Bradford, Bradford BD7 1DP, UK}

\date{}

\abstract{}{In this work we study time-dependent precipitation of an electron beam injected into a flaring atmosphere with a converging magnetic field by considering collisional and Ohmic losses with anisotropic scattering and pitch angle diffusion. Two injection regimes are investigated: short impulse and stationary injection. The effects of converging magnetic fields with different spatial profiles are compared and the energy deposition produced by the precipitating electrons at different depths and regimes is calculated. }{The time dependent Fokker-Planck equation for electron distribution in depth, energy and pitch angle was solved numerically by using the summary approximation method.}{It was found that steady state injection is established for beam electrons at 0.07-0.2 seconds after the injection onset depending on the initial beam parameters. Energy deposition by a stationary beam is strongly dependent on a self-induced electric field but less on a magnetic field convergence. Energy depositions by short electron impulses are found to be insensitive to the self-induced electric field but are strongly affected by a magnetic convergence. Short beam impulses are shown to produce sharp asymmetric hard X-ray bursts within a millisecond timescale often observed in solar flares.}{}


\maketitle

\section{Introduction}

Observations of solar flares in hard X-rays provide vital information about scenarios, in which accelerated electrons gain and deposit their energy into flaring atmospheres. In recent years the theory describing the generation of bremsstrahlung emission has been significantly progressed in many directions by improving the mechanisms for emitting this radiation, e.g. considering relativistic bremsstrahlung cross-sections \citep{Kontar06}, taking into account various aspects of the photospheric albedo effects while deriving mean electron spectra from the observed bremsstrahlung photon spectra \citep{Kontar06}. On the other hand, substantial improvements were also achieved in the solutions of the direct problem of electron precipitation into a flaring atmosphere by taking into account different mechanisms of electron energy losses: Coulomb collisions \citep{Brown71, Brown00} combined with the deceleration by the self-induced electric field \citep{Zharkova95, Zharkova06} in flaring atmospheres with strong temperature and density gradients derived from the hydro-dynamic solutions \citep{Somov81, Somov82, Nagai84, Fisher85}.

Substantial progress in the quantitative interpretation of hard X-ray emission has been made in recent years by using high temporal and spatial resolution observations carried out by the RHESSI payload \citep{Lin03}. The latter provides the locations and shapes of hard X-ray sources on the solar disk, their temporal variations and energy spectra evolution during the flare duration \citep{Holman03, Krucker08}. These observations are often accompanied by other observations (in microwaves (MW), EUV and optical ranges) which reveal a very close temporal correlation between HXR and MW, UV and even optical emission \citep[see for example,][]{Kundu04, Fletcher07, Grechnev08}. This highlighted a further need for the improvements of electron transport models, which can simultaneously account both temporarily and spatially for all these types of emission.

The RHESSI observations of double power law energy spectra with flattening towards lower photon energies \citep{Holman03}, which leads to the soft-hard-soft temporal pattern of the photon spectra indices below $35 \units{keV}$ \citep{Grigis06}, highlighted their role in the formation of the self-induced electric field \citep{Zharkova06}. The authors considered a stationary beam injection and naturally reproduced such a spectral flattening by electron deceleration in the electric field, induced by beam electrons themselves. The flattening was shown to be proportional to the initial energy flux of beam electrons and their spectral indices \citep{Zharkova06}. Then the soft-hard-soft pattern in photon spectra above can be easily reproduced by a triangular increase and decrease of the beam energy flux in the time interval of a few seconds, which was often observed both by RHESSI \citep{Lin03}, and by the previous SMM mission \citep{Kane80}. Furthermore, numerous observations of solar flares by SMM, TRACE and RHESSI suggest that the areas of flaring loops decrease and, thus, their magnetic fields increase with the depth of the solar atmosphere \citep{Lang93, Brosius06, Kontar08}. This increase of magnetic field can act as a magnetic mirror for the precipitating electrons forcing them to return back to the source in the corona, in addition to self-induced electric field.

In the present paper we propose two models of the magnetic field variations with depth. One model is a fitting to the measurements of the magnetic field in the corona \citep{Brosius06} and chromosphere \citep{Kontar08}, another shows the exponential increase of magnetic field from the corona to the upper chromosphere while remaining a constant in the lower chromosphere. The outcome is compared with the two other models proposed earlier: the first one \citep{Leach81} where  the authors assumed that the magnetic column depth scale $\partial\ln B / \partial s$ is constant implying the exponential magnetic field increase with a column depth and the second one \citep{McClements92} considering a parabolic increase of the magnetic field with a linear depth.

Also the temporal intervals of impulsive increases of HXR emission vary from very short (tens of milliseconds \citep{Kiplinger83, Charikov04}) to tens of minutes often observed by RHESSI \citep{Holman03}. This encouraged us to revise the electron transport models and to consider solutions of a time-dependent Fokker-Planck equation for different timescales of beam injection (milliseconds, seconds and minutes). The electron transport, in turn, can slow down also by anisotropic scattering of beam electrons in this self-induced electric field enhanced by their magnetic mirroring in converging magnetic loops. The further delay can be caused by the particle diffusion in pitch angles and energy which can significantly extend the electron transport time into deeper atmospheric layers where they are fully thermalised.

We also apply the time dependent Fokker-Planck equation in order to compare the solutions for electron precipitation for stationary and impulsive injection and their effect on resulting hard X-ray emission and ambient plasma heating for different parameters of beam electrons. We also investigate these Fokker-Planck solutions for the different models of a converging magnetic field by taking into account all the mechanisms of energy loss (collisions, Ohmic losses) and anisotropic scattering but without diffusion in energy.

The problem is formulated in Sect.~\ref{sec:problem} and the method of solution is described in Sect.~\ref{sec:method}. The stationary injection into flaring atmosphere with different magnetic field convergence and collisional plus Ohmic losses with anisotropic scattering is considered in Sect.~\ref{sec:stationar} and the impulsive injection for short timescales below tens milliseconds is considered in Sect.~\ref{sec:impulse}. The discussion and conclusions are drawn in Sect.~\ref{sec:concl}.

\section{Problem formulation} \label{sec:problem}

\subsection{The Fokker-Planck equation} \label{sec:equations}

We consider a one-dimensional beam of high energy electrons, that is injected into the solar atmosphere. The beam electron velocity distribution $f$, as a function of time $t$, depth $l$, velocity $V$ and pitch angle between the velocity and the magnetic field $\theta$, can be found by solving the Fokker-Planck equation \citep{Diakonov88, Zharkova95}:
\begin{equation}\label{eq:fok_pl_dim}
\begin{split}
    &\frac{\partial f}{\partial t} + V \cos\theta \frac{\partial f}{\partial l} -
    \frac{e\mathcal E}{m_\mathrm{e}} \cos\theta \frac{\partial f}{\partial V} -
    \frac{e\mathcal E}{m_\mathrm{e} V} \sin^2\theta \frac{\partial f}{\partial \cos\theta} = \\
    &\frac{1}{V^2} \frac{\partial}{\partial V} \left( \nu V^3 f \right) +
    \nu \frac{\partial}{\partial \cos\theta} \left( \sin^2\theta \frac{\partial f}{\partial \cos\theta} \right) + \\
    &\frac{V\sin^2\theta}{2} \frac{\partial\ln B}{\partial l} \frac{\partial f}{\partial \cos\theta},
\end{split}
\end{equation}
where the collisional rate $\nu$ is given by
\begin{equation}
    \nu = n_\mathrm{p}(l) \frac{2 \pi e^4 \ln\Lambda}{m_\mathrm{e}^2 V^3},
\end{equation}
$\mathcal E$ is the self-induced electric field, $B$ is the background magnetic field, $n_\mathrm{p}(l)$ is the density of the ambient plasma, $\ln\Lambda$ is the Coulomb logarithm, $e$ and $m_\mathrm{e}$ are the electron charge and mass respectively. In our study we assume that the Coulomb logarithm is constant: $\ln\Lambda\approx20$.

Let us introduce the following dimensionless variables:
\begin{align}
    \tau &= t \frac{2\pi e^4 n_0\ln\Lambda}{\sqrt{2m_\mathrm{e}}E_0^{3/2}},\\
    s &= \xi \frac{\pi e^4 \ln\Lambda}{E_0^2},\\
    z &= \frac{E}{E_0} = \frac{m_\mathrm{e} V^2}{2 E_0},\\
    \mu &= \cos\theta,\\
    \varepsilon &= \mathcal E \frac{E_0}{2\pi e^3 n_0\ln\Lambda},\\
    n &= \frac{n_\mathrm{p}}{n_0}
\end{align}
where $\xi$ is the column depth:
\begin{equation}
    \xi = \int\limits_0^l n_\mathrm{p}\left(l^{\prime}\right)dl^{\prime},
\end{equation}
$E_0=12\units{keV}$ is the lower cut-off energy and $n_0 = 10^{10}\percmcub$. Eq.~(\ref{eq:fok_pl_dim}) in dimensionless variables takes the form
\begin{equation}\label{eq:fok_pl}
\begin{split}
    &\frac{\partial f}{\partial\tau} + n\sqrt{z}\mu\frac{\partial f}{\partial s} - 2\varepsilon\mu\sqrt{z}\frac{\partial f}{\partial z} -
    \varepsilon\frac{1-\mu^{2}}{\sqrt{z}}\frac{\partial f}{\partial\mu} = \\
    &n\frac{1}{\sqrt{z}}\frac{\partial f}{\partial z} +
    n\frac{1-\mu^{2}}{2z^{3/2}}\frac{\partial^{2}f}{\partial\mu^{2}} - n\frac{\mu}{z^{3/2}}\frac{\partial f}{\partial\mu} + \\
    &n\frac{\left(1-\mu^{2}\right)\sqrt{z}}{2}\alpha_\mathrm{B}\frac{\partial f}{\partial\mu},
\end{split}
\end{equation}
where $\alpha_\mathrm{B}$ is the magnetic convergence parameter defined as
\begin{equation}
    \alpha_\mathrm{B} = \frac{\partial\ln B}{\partial s}.
\end{equation}

\begin{figure}
  \centering
  \resizebox{\hsize}{!}{\includegraphics{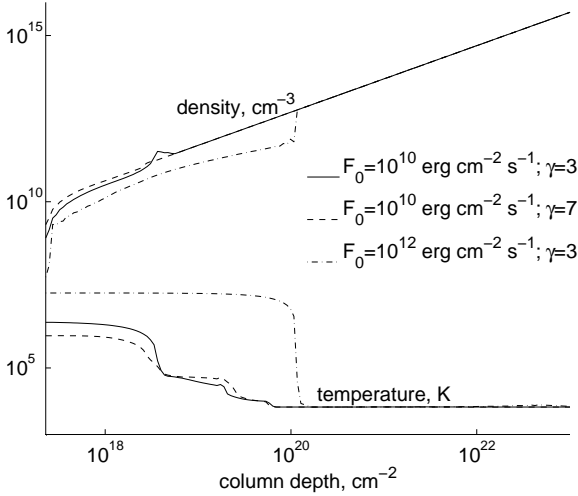}}
  \caption{Density and temperature of the ambient plasma calculated by the hydro-dynamic model \citep{Zharkova07} for different beam energy flux, $F_0$, and power law index, $\gamma$, of the beam electron distribution (see Eq.~(\ref{eq:initdistr})).}
  \label{fig:hd_model}
\end{figure}

As it was shown by \citet{Oord90} after a few collisional times from the start of the beam injection the background plasma fully compensates the charge and current deposited by the beam. In our model we assume that the beam current is always compensated, i.e. $j_b + j_p = 0$, where $j_b$ and $j_p$ are the beam and plasma currents respectively. Note, that beam current $j_b$ includes electrons with $\mu<0$, i.e. those that moves back to source, which can significantly decrease the compensation current $j_p$. Then, by applying the Ohm's law for $j_p$ and taking into account that $j_p = -j_b$ we can find the self-induced electric field \citep{McClements92},
\begin{equation}\label{eq:el_field}
    \varepsilon\left(\tau,s\right) = \frac{1}{\sigma\left(s\right)}\int\limits_{z_\mathrm{min}}^{z_\mathrm{max}} dz \int\limits_{-1}^{1} d\mu z \mu f\left(\tau,s,z,\mu\right),
\end{equation}
where the dimensionless conductivity is
\begin{equation} \label{eq:conduct}
    \sigma\left(s\right) = 1.97\sqrt{2\pi}\frac{3}{4}\frac{n_0\left(kT\left(s\right)\right)^{3/2}}{\sqrt{m_\mathrm{e}}F_0},
\end{equation}
and $F_0=10^{10} \fluxunit$ is the normalisation factor for the energy flux of the beam. The ambient plasma is assumed to be preheated and its density, $n$, and temperature, $T$, as functions of the column depth, $s$, are calculated using the hydro-dynamic model \citep{Zharkova07}. The profiles $n(s)$ and $T(s)$ for different beam parameters are shown in Fig.~\ref{fig:hd_model}. These profiles do not change in time, since the thermal conduction processes have much longer time scales than the precipitation processes studied here.

The electron beam as well as the plasma return current can be a subject of various plasma instabilities. For example, \citet{Zharkova92} have shown that a beam with energy flux higher than $10^{11} \fluxunit$ can be unstable with respect to the ion-sound waves. However, such instabilities are out of scope of the current paper and will be considered in forthcoming studies.

\subsection{Initial and boundary conditions}

There are no beam electrons before the injection starts, thus, the initial condition is:
\begin{equation}
    f\left(\tau=0,s,z,\mu\right) = 0.
\end{equation}

The boundary condition at $s=s_\mathrm{min}=2.08\tento{-3}$ (or $2.29 \tento{17}\percmsq$) corresponds to the injected beam distribution
\begin{equation} \label{eq:initdistr}
    f\left(\tau,s=s_\mathrm{min},z,\mu>0\right) = f_\mathrm{n} \psi\left(\tau\right) \frac{z^{\delta-1}}{z^{\delta+\gamma}+1} \exp\left(-\frac{\left(1-\mu\right)^2}{\Delta\mu^2}\right),
\end{equation}
where $\Delta\mu$ is the initial pitch angle dispersion and $\psi\left(\tau\right)$ determines the time variation of the beam. If the energy is much larger than the lower cut-off energy, $z \gg 1$, the distribution is power law with index $-\gamma-1$, thus, the flux spectrum ($\sim zf$) is power law with index $\gamma$. In the opposite case, $z \ll 1$, the distribution is power law with index $\delta-1$. The low energy index $\delta$ is chosen to be $10$ \citep[see, e.g.,][]{Zharkova05}, while for the high energy index two values, $3$ and $7$, are considered. $f_\mathrm{n}$ is the normalisation coefficient, which is chosen so that the energy flux of the injecting electron beam,
\begin{equation}
    F\left(s=s_\mathrm{min}\right) = F_0 \int\limits_{z_\mathrm{min}}^{z_\mathrm{max}} dz \int\limits_{-1}^{1} d\mu z^2 \mu f\left(s=s_\mathrm{min},z,\mu\right),
\end{equation}
is equal to some preset value $F_\mathrm{top}$, where $F_0=10^{10} \fluxunit$ is the normalisation factor of the energy flux. At large depth, $s=s_\mathrm{max}=9.17\tento{2}$ (or $1.01 \tento{23}\percmsq$), the number of electrons in the beam is assumed to be negligibly small, thus the corresponding boundary condition is
\begin{equation}
    f\left(\tau,s=s_\mathrm{max},z,\mu<0\right) = 0
\end{equation}

The distribution function is calculated in the following range of energies: $z_\mathrm{min} \leq z \leq z_\mathrm{max}$, where $z_\mathrm{min}=0.1$ (or $1.2 \units{keV}$) and $z_\mathrm{max}=100$ (or $1.2 \units{MeV}$). The boundary conditions on energy are
\begin{align}
    \frac{\partial f\left(\tau,s,z=z_\mathrm{min},\mu\right)}{\partial z} &= 0, \\
    \frac{\partial f\left(\tau,s,z=z_\mathrm{max},\mu\right)}{\partial z} &= 0.
\end{align}

The boundary conditions on pitch angle are \citep{McClements90}
\begin{align}
    \frac{\partial f\left(\tau,s,z,\mu=1\right)}{\partial \mu} &= 0,\\
    \frac{\partial f\left(\tau,s,z,\mu=-1\right)}{\partial \mu} &= 0.
\end{align}

\subsection{Integral characteristics of the electron distribution in the beam}

In the following sections we will numerically solve Eq.~(\ref{eq:fok_pl}) and calculate the following quantities for the electron beam: beam density (in $\percmcub$),
\begin{equation}
    n_\mathrm{b}\left(\tau,s\right) = F_0 \sqrt{\frac{m_\mathrm{e}}{2 E_0^3}} \int\limits_{z_\mathrm{min}}^{z_\mathrm{max}} dz \int\limits_{-1}^{1} d\mu \sqrt{z} A\left(s\right) f\left(\tau,s,z,\mu\right),
\end{equation}
differential particle flux spectrum (in $\fluxspunit$),
\begin{equation}
    \mathcal{F}_\mathrm{n} \left(\tau,s,z\right) = \frac{F_0}{2 E_0^2} \int\limits_{-1}^{1} d\mu z A\left(s\right) f\left(\tau,s,z,\mu\right),
\end{equation}
mean particle flux spectrum (in $\fluxspunit$) \citep{Brown03},
\begin{equation}
    \left\langle \mathcal{F}_\mathrm{n} \right\rangle \left(\tau,z\right) = \frac{F_0}{E_0^2} \frac{\int\limits_{s_\mathrm{min}}^{s_\mathrm{max}} ds \int\limits_{-1}^{1} d\mu  n^{-1}\left(s\right) A\left(s\right) z f\left(\tau,s,z,\mu\right)}{2 \int\limits_{s_\mathrm{min}}^{s_\mathrm{max}} n^{-1}\left(s\right) ds},
\end{equation}
angle distribution (in arbitrary units),
\begin{equation}
    \frac{dN_\mathrm{b}\left(\tau,\mu\right)}{d\mu} = \int\limits_{s_\mathrm{min}}^{s_\mathrm{max}} ds \int\limits_{z_\mathrm{min}}^{z_\mathrm{max}} dz n^{-1}\left(s\right) A\left(s\right) \sqrt{z} f\left(\tau,s,z,\mu\right),
\end{equation}
and energy deposition (or heating function) of the beam (in $\heatunit$),
\begin{equation}
    I\left(\tau,s\right) = \frac{F_0 n_0}{E_0} n\left(s\right)  A\left(s\right) \int\limits_{z_\mathrm{min}}^{z_\mathrm{max}} dz \int\limits_{-1}^{1} d\mu \left(-\frac{dz}{ds}\right) \mu z f\left(\tau,s,z,\mu\right),
\end{equation}
where $dz/ds$ is the electron's energy losses with depth, which can be estimated as \citep{Emslie80}
\begin{equation}
    \frac{dz}{ds} = \left( \frac{dz}{ds}\right)_\mathrm{c} + \left(\frac{dz}{ds}\right)_\mathrm{r} = -\frac{1}{\mu z} - 2\frac{\varepsilon}{n},
\end{equation}
where two terms represent the collisional and Ohmic energy losses respectively. Coefficient $A\left(s\right) = B_0/B\left(s\right)$ takes into account the variation of the magnetic tube cross-section.

\section{Summary approximation method} \label{sec:method}

Let us combine the relative derivatives and rewrite the Fokker-Planck equation~(\ref{eq:fok_pl}) in the following form
\begin{equation}\label{eq:fok_pl1}
\begin{split}
    &\frac{\partial f}{\partial\tau} = -n\sqrt{z}\mu \frac{\partial f}{\partial s} +
    \left( 2\varepsilon\mu\sqrt{z} + n\frac{1}{\sqrt{z}} \right) \frac{\partial f}{\partial z} + \\
    &\left( \varepsilon\frac{1-\mu^{2}}{\sqrt{z}} - n\frac{\mu}{z^{3/2}} +
    n\frac{\left(1-\mu^{2}\right)\sqrt{z}}{2}\alpha_\mathrm{B} \right) \frac{\partial f}{\partial\mu} + \\
    &n\frac{1-\mu^{2}}{2z^{3/2}} \frac{\partial^{2}f}{\partial\mu^{2}} =
    \phi_s \frac{\partial f}{\partial s} + \phi_z \frac{\partial f}{\partial z} +
    \phi_{\mu} \frac{\partial f}{\partial \mu} + \phi_{2\mu} \frac{\partial^2 f}{\partial \mu^2},
\end{split}
\end{equation}

Eq.~(\ref{eq:fok_pl1}) is solved numerically by using the summary approximation method \citep{Samarskii01}. This method allows us to study time dependent Fokker-Planck equation and it is different from the one used by \citet{Zharkova05aa} to solve the stationary problem. According to the summary approximation method the four-dimensional problem is reduced to a chain of three two-dimensional problems. This is done by considering the three-dimensional differential operator at the right hand side of Eq.~\ref{eq:fok_pl} as a sum of one-dimensional operators, each acting on the distribution function separately during one third of the time step. On each time substep the distribution function is calculated implicitly, hence, the numerical scheme is
\begin{align}
    f^{\tau + \frac{1}{3}\Delta \tau} - f^{\tau} &= \Delta \tau \phi_s L_s f^{\tau + \frac{1}{3}\Delta \tau} \label{eq:scheme1}\\
    f^{\tau + \frac{2}{3}\Delta \tau} - f^{\tau + \frac{1}{3}\Delta \tau} &= \Delta \tau \phi_z L_z f^{\tau + \frac{2}{3}\Delta \tau} \label{eq:scheme2}\\
    f^{\tau + \Delta \tau} - f^{\tau + \frac{2}{3}\Delta \tau} &= \Delta \tau \left( \phi_{\mu} L_{\mu} + \phi_{2\mu} L_{2\mu} \right) f^{\tau + \Delta \tau} \label{eq:scheme3}
\end{align}
where $L_{\alpha}$ are the finite difference operators that approximate the first order differential operators $\partial / \partial \alpha$. If the coefficient $\phi_{\alpha}$ is positive then the right difference scheme is used, i.e. $L_{\alpha} f = \left( f^{\alpha+\Delta \alpha} - f^{\alpha} \right)/\Delta \alpha$, otherwise the left scheme is used, i.e. $L_{\alpha} f = \left( f^{\alpha} - f^{\alpha-\Delta \alpha} \right)/\Delta \alpha$. The $L_{2\mu} f = \left( f^{\mu+\Delta \mu} - 2f^{\mu} + f^{\mu-\Delta \mu} \right)/\Delta \alpha^2$ is the central difference that approximates the second order derivative $\partial^2 f / \partial \mu^2$. The computational grid has 200 nodes in the $s$ dimension, 50 nodes in the $z$ dimension and 30 nodes in the $\mu$ dimension. The nodes are distributed logarithmically in the $s$ and $z$ dimensions and linearly in the $\mu$ dimension.

Eq.~(\ref{eq:scheme1}) together with boundary conditions forms a set of linear equations, which, after being solved, gives $f^{\tau + \frac{1}{3}\Delta \tau}$ from known $f^{\tau}$. Distribution function $f^{\tau + \frac{1}{3}\Delta \tau}$ is then used in Eq.~(\ref{eq:scheme2}) to obtain $f^{\tau + \frac{2}{3}\Delta \tau}$. Finally, from Eq.~(\ref{eq:scheme3}) we obtain $f^{\tau + \Delta \tau}$ which is, in turn, used in Eq.~(\ref{eq:scheme1}) on the next step.

Electric field $\varepsilon$ is calculated on each time step according to Eq.~(\ref{eq:el_field}) where the distribution function is taken from the previous step. Thus, the numerical scheme is not fully implicit. This means that, in order to avoid numerical instability, the time step, $\Delta \tau$, must be shorter than some critical value $\Delta \tau_c$. In practice, the time step was determined by the trial-and-error method. For example, for the energy flux $10^{10} \fluxunit$ the time step is $1.7\tento{-4}\units{s}$. It was found that when the energy flux is increased, the time step need to be decreased proportionally to keep the numerical scheme stable.

\section{Stationary injection} \label{sec:stationar}

\begin{figure}
  \centering
  \subfloat[]{
    \resizebox{\hsize}{!}{\includegraphics{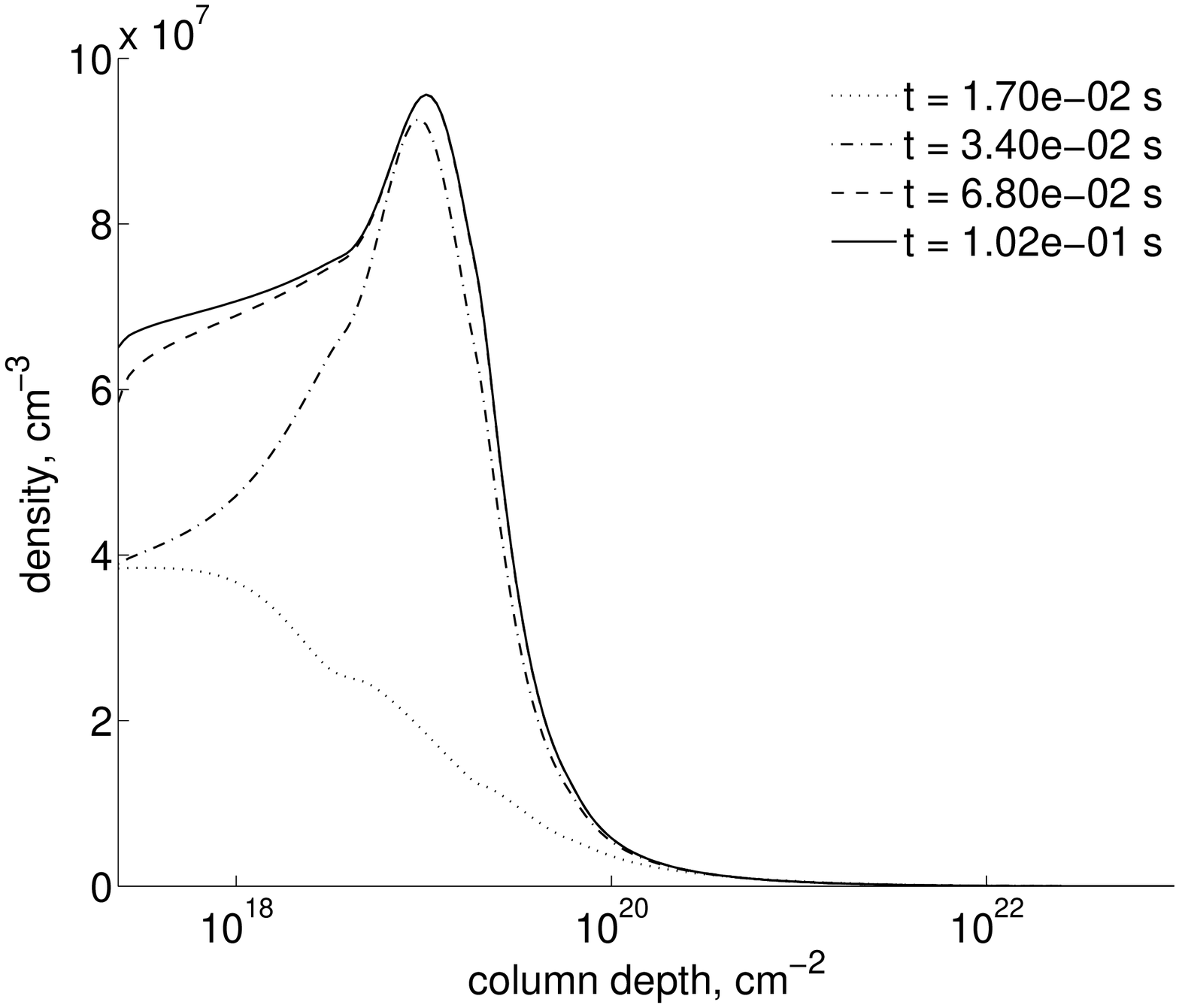}}
    \label{fig:relax_f}}\quad
  \subfloat[]{
    \resizebox{\hsize}{!}{\includegraphics{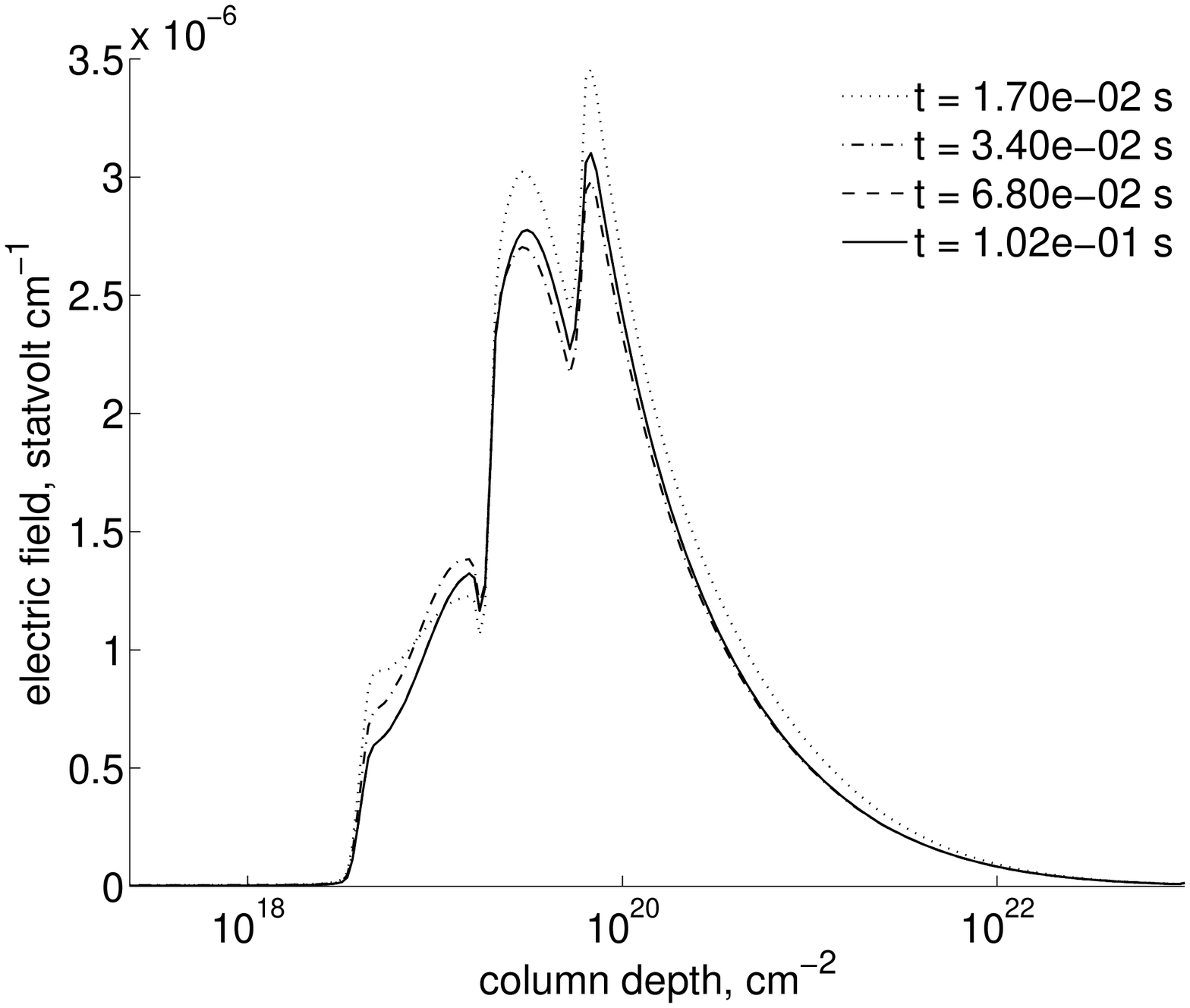}}
    \label{fig:relax_E}}
  \caption{Electron density (a) and self-induced electric field (b) profiles, where $t$ is time passed after the injection is "turned on". Collisions and electric field are taken into account. The beam parameters, see Eq.~(\ref{eq:initdistr}), are $\gamma = 3$, $F_\mathrm{top} = 10^{10} \fluxunit$ and $\Delta\mu = 0.2$.}
  \label{fig:relax}
\end{figure}

Let us first recapture the major simulation results for the case of a stationary (steady) injection which helps to understand the physical process of energy losses by beam electrons during their precipitation. In our previous paper \citet{Zharkova09a} we studied hard X-ray emission and polarisation produced by a steady beam while in the current paper we focus more on the comparison of electron precipitation results obtained for different models of magnetic convergence and on the energy deposition at various depths and times by electron beams with different parameters.

The electron injection starts at $t=0$ and the simulation continues until the stationary state is reached. The initial spectral index of the beam is chosen to be $\gamma = 3$, the energy flux at the top boundary is $F_\mathrm{top} = 10^{10} \fluxunit$ and the initial angle dispersion is $\Delta\mu = 0.2$.

\subsection{Relaxation time to a steady injection} \label{sec:relax}

Let us first show how the system relaxes to the stationary state. Fig.~\ref{fig:relax} shows the profiles of the electric field and beam density at different times. It is seen that the electric field relaxes somewhat faster than the density. The relaxation time $t_\mathrm{r}$, after which the system becomes stationary, can be estimated as $\sim 0.07\units{s}$. As it was stated in Sec.~\ref{sec:equations} the self-induced electric field is calculated under the assumption that the beam current changes on time scales longer than the plasma collisional time. Since the collisional time in the corona is about $0.05 \units{s}$ and it decreases essentially with depth we may conclude that this assumption is correct.

It was found that this relaxation time is longer for a stronger beam, which is the result of a smaller density of the ambient plasma (Fig.~\ref{fig:hd_model}) and, hence, a larger linear depth for the same column depth. For example, the depth $10^{20} \percmsq$ corresponds to $2\tento{8}\units{cm}$ and $11\tento{8}\units{cm}$ for the beam energy fluxes $10^{10}\fluxunit$ and $10^{12}\fluxunit$ respectively. This leads to a longer relaxation time for the atmosphere preheated by a stronger beam, e.g. for the beam energy flux of $10^{12}\fluxunit$ the relaxation time is found to be $\sim 0.2 \units{s}$.

\subsection{Effects of collisions and electric field ($\alpha_\mathrm{B}=0$)}

\begin{figure}
  \centering
  \resizebox{\hsize}{!}{\includegraphics{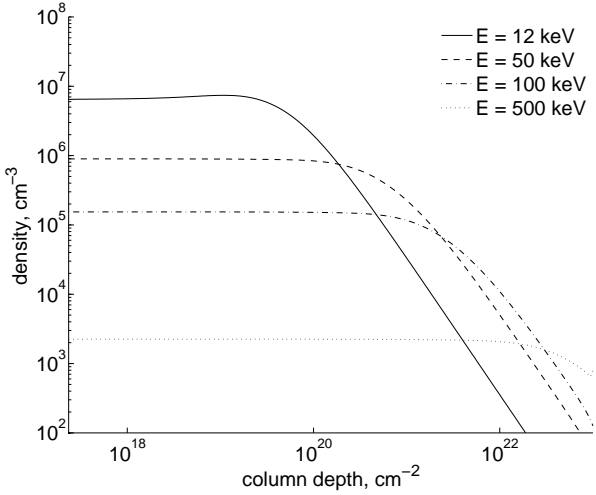}}
  \caption{Beam electron density in different energy bands, if only collisions are taken into account. Beam parameters are the same as in Fig.~\ref{fig:relax}.}
  \label{fig:loss-const-Bc0-noE}
\end{figure}

The local maximum, which appears on the density profile at the depth of about $\sim 2 \tento{19} \percmsq$ (Fig.~\ref{fig:relax_f}), is caused by the beam deceleration while the flux of electrons remains nearly constant. After this depth most of the electrons leave the distribution (thermalise) by reducing their energy below the energy $z_\mathrm{min}$, and the density rapidly decreases at the depth of about $\sim 5 \tento{19} \percmsq$ (Fig.~\ref{fig:relax_f}). This corresponds to the stopping depth for the electrons with energies close to the lower cut-off energy ($12 \units{keV}$). The electrons with higher energies can travel deeper (see Fig.~\ref{fig:loss-const-Bc0-noE}). In particular, it can be seen that electrons with energies $>500 \units{keV}$ can travel down to the photosphere almost without any energy losses. Fig.~\ref{fig:loss-const-Bc0-noE} confirms the results obtained by \citet{Zharkova06} (see their Tab. 1).

The depth profile of the electric field (Fig.~\ref{fig:relax_E}) has some noticeable variations at the column depths of about $4 \tento{18} \percmsq$, $2 \tento{19} \percmsq$ and $6 \tento{19} \percmsq$. These features are located at the same depths where the temperature profile (Fig.~\ref{fig:hd_model}) has sharp decreases leading to the sharp decreases of the plasma conductivity given by Eq.~(\ref{eq:conduct}). However, the electron density fluctuation associated (through the Gauss's Law) with such an electric field profile is found to be of the order of $10^{-4} \percmcub$ and can be neglected.

\begin{figure*}
  \centering
  \subfloat[collisions, $\mu>0$]{
    \includegraphics[width=0.45\textwidth]{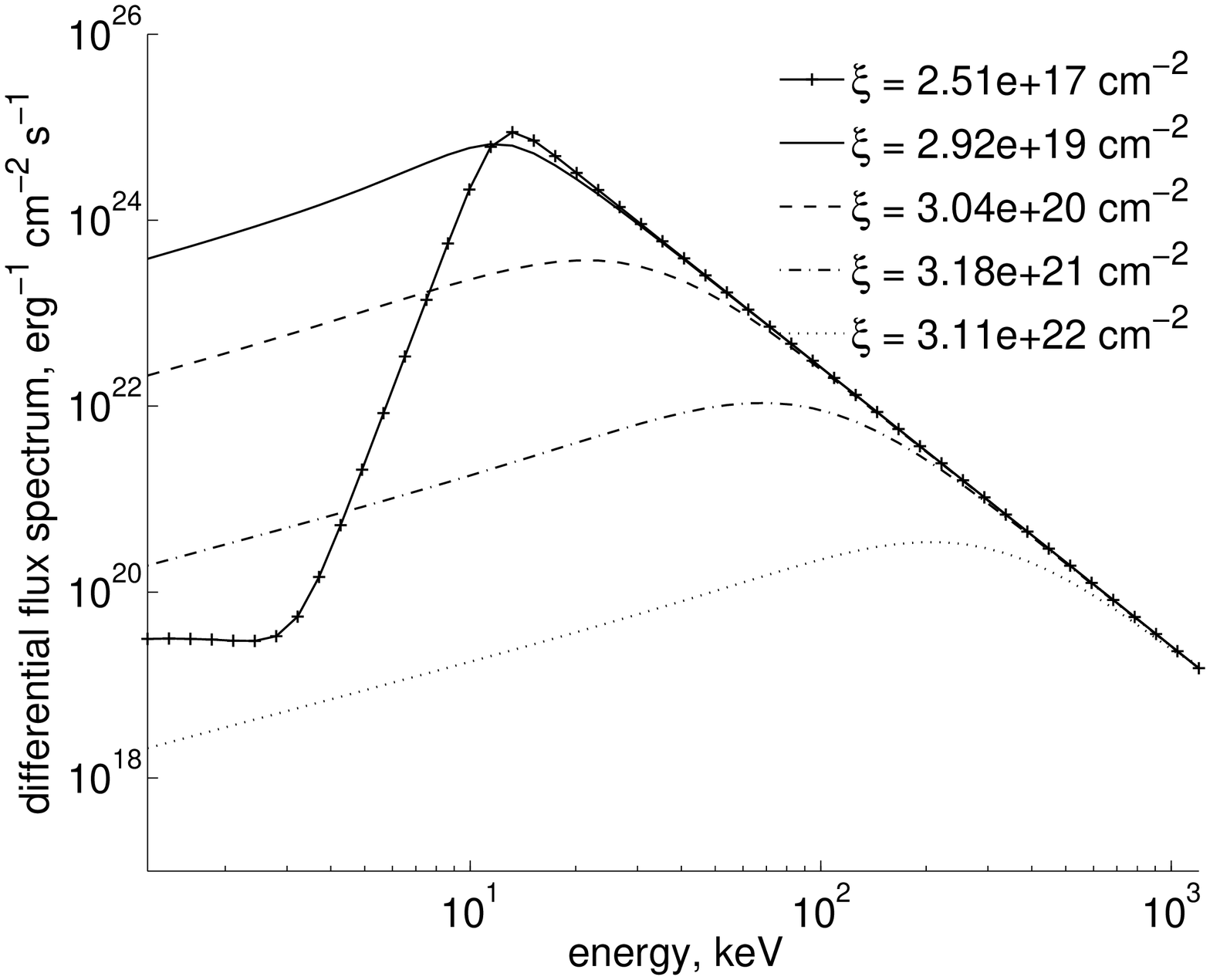}
    \label{fig:difp-const-Bc0-noE}}\quad
  \subfloat[collisions and electric field, $\mu>0$]{
    \includegraphics[width=0.45\textwidth]{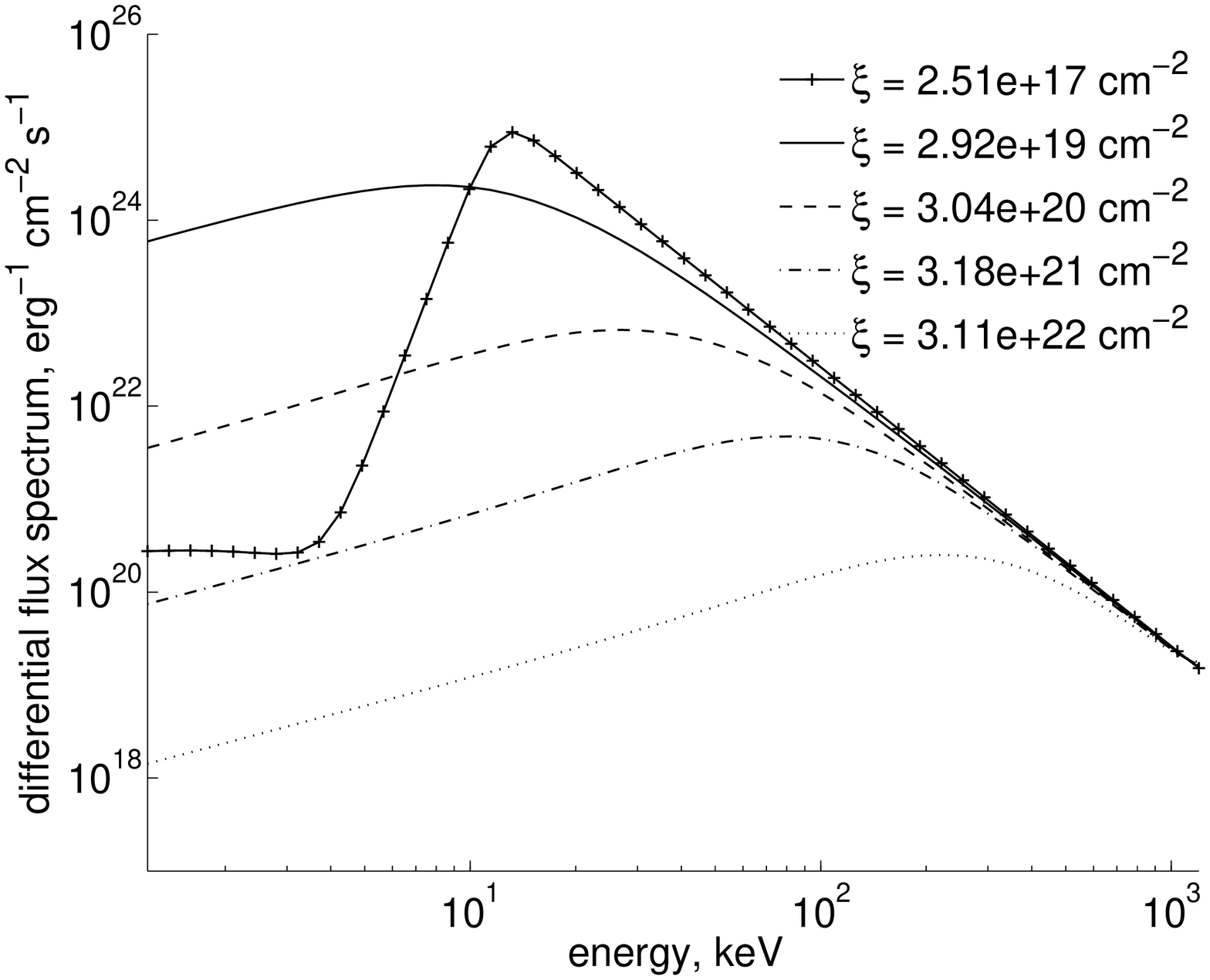}
    \label{fig:difp-const-Bc0}} \\
  \subfloat[collisions, $\mu<0$]{
    \includegraphics[width=0.45\textwidth]{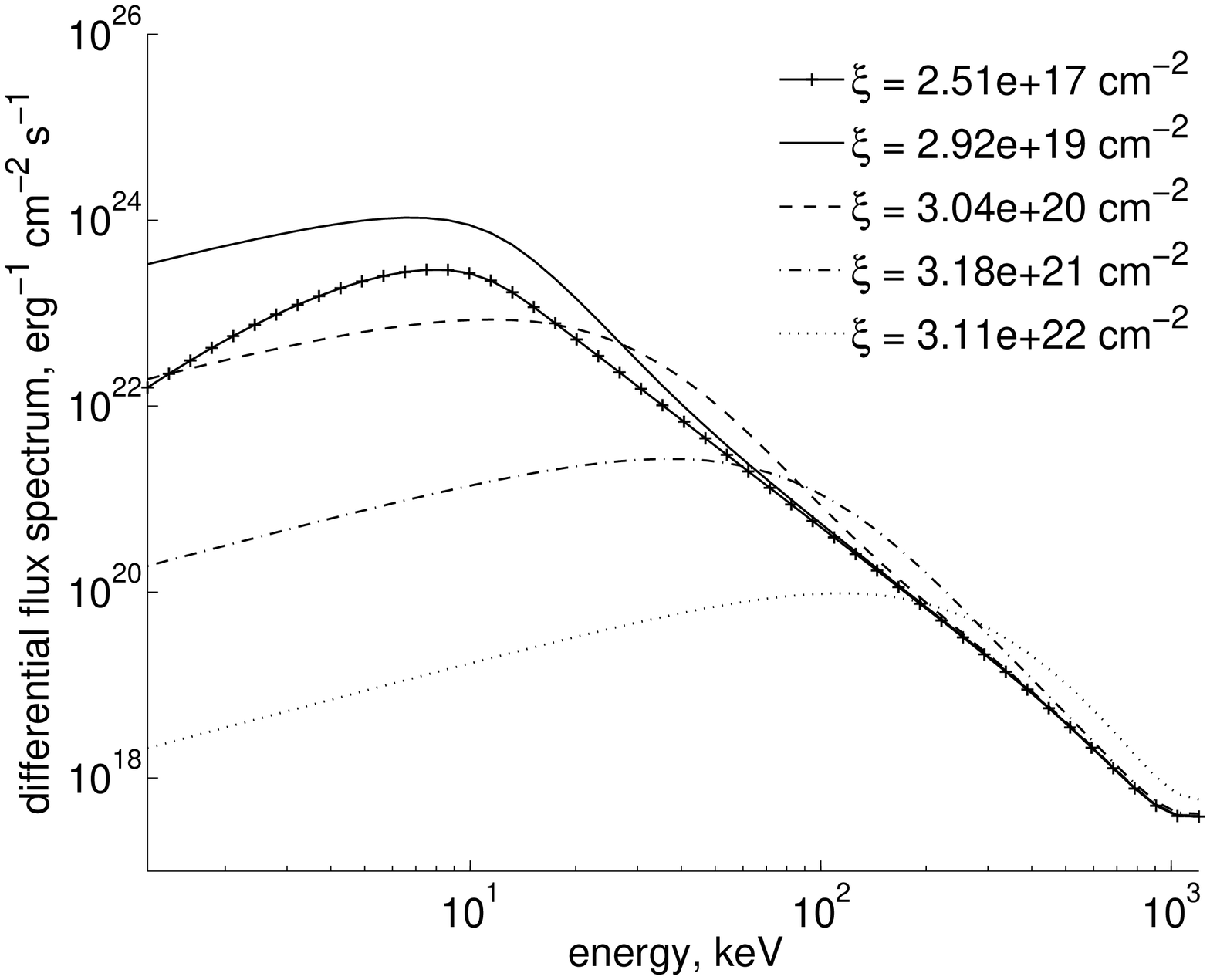}
    \label{fig:difm-const-Bc0-noE}}\quad
  \subfloat[collisions and electric field, $\mu<0$]{
    \includegraphics[width=0.45\textwidth]{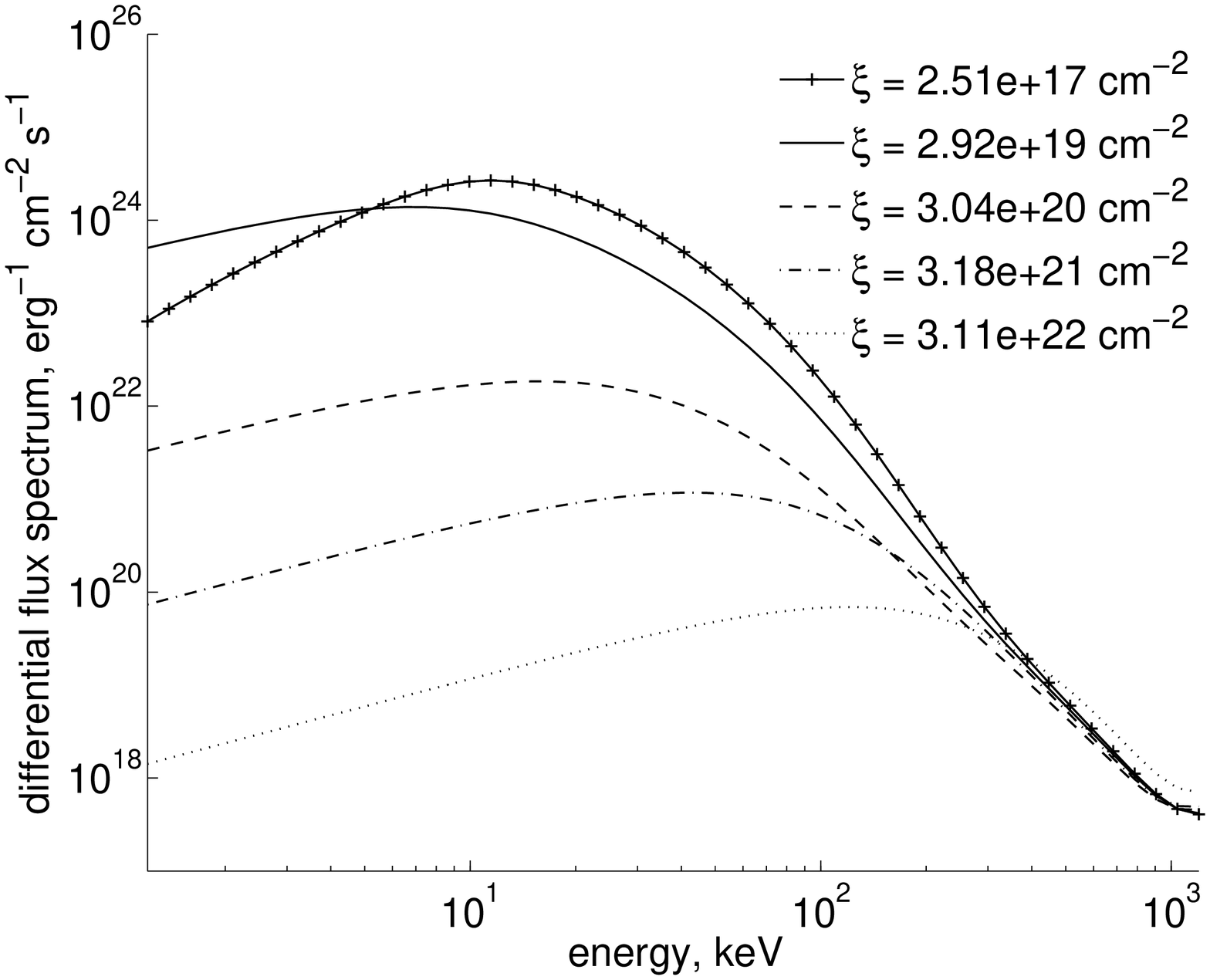}
    \label{fig:difm-const-Bc0}}
  \caption{Differential flux spectra of the beam electrons integrated over the positive and negative pitch angles. Beam parameters are the same as in Fig.~\ref{fig:relax}.}
  \label{fig:dif-const-Bc0-all}
\end{figure*}

In Fig.~\ref{fig:dif-const-Bc0-all} we plot the differential flux spectra at different depths for the forward ($\mu>0$) and backward ($\mu<0$) moving electrons. It can be seen that the self-induced electric field does not change the spectra of the downward ($\mu>0$) moving electrons but essentially affects the spectra of the upward ($\mu<0$) moving ones. Since the pitch angle diffusion due to the electric field is more effective for the lower energy electrons, the spectra of the returned electrons is enhanced at low and mid energy (Fig.~\ref{fig:difm-const-Bc0}).

\begin{figure}
  \centering
  \subfloat[]{
    \resizebox{\hsize}{!}{\includegraphics{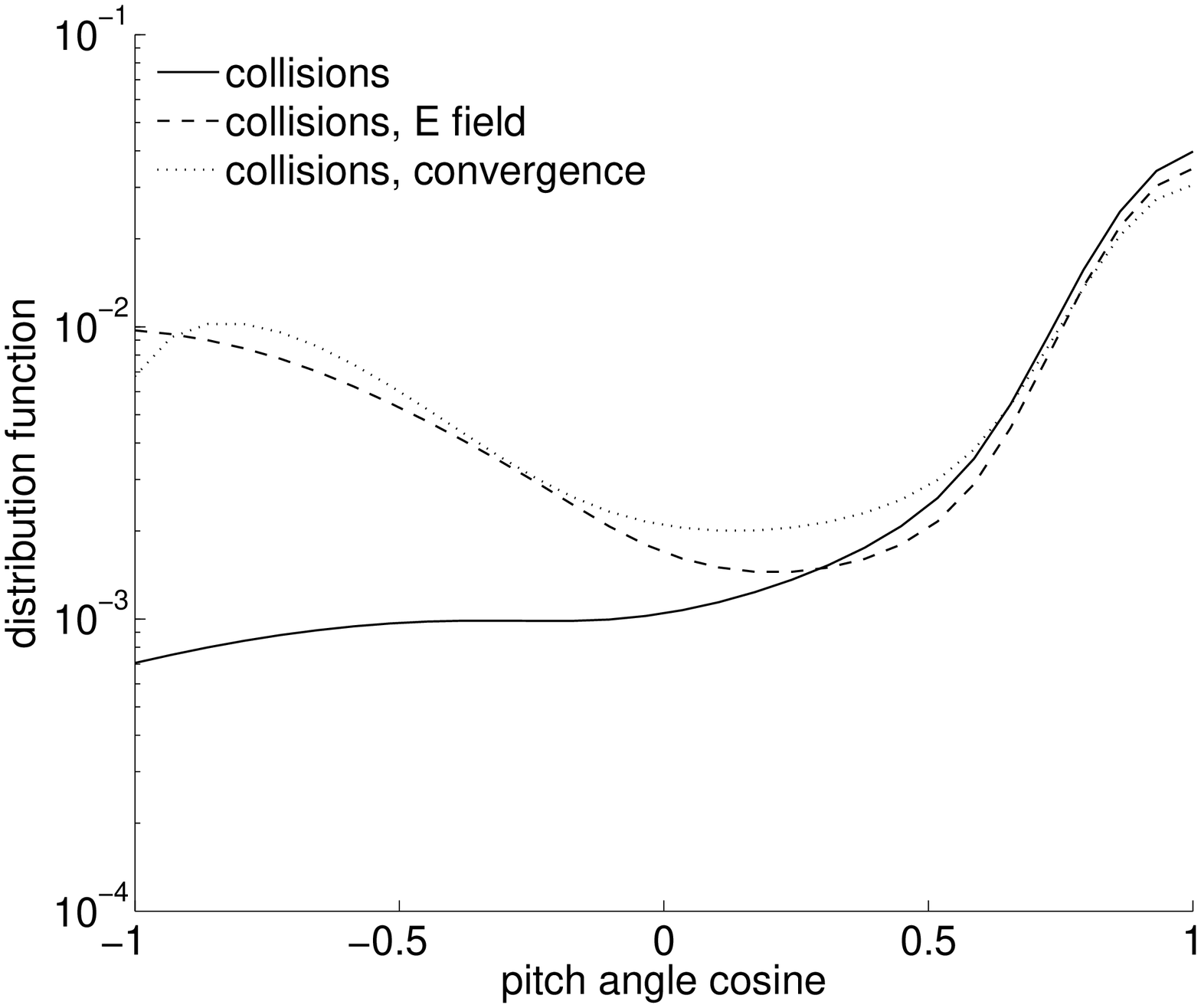}}
    \label{fig:mu-const-all}}\quad
  \subfloat[]{
    \resizebox{\hsize}{!}{\includegraphics{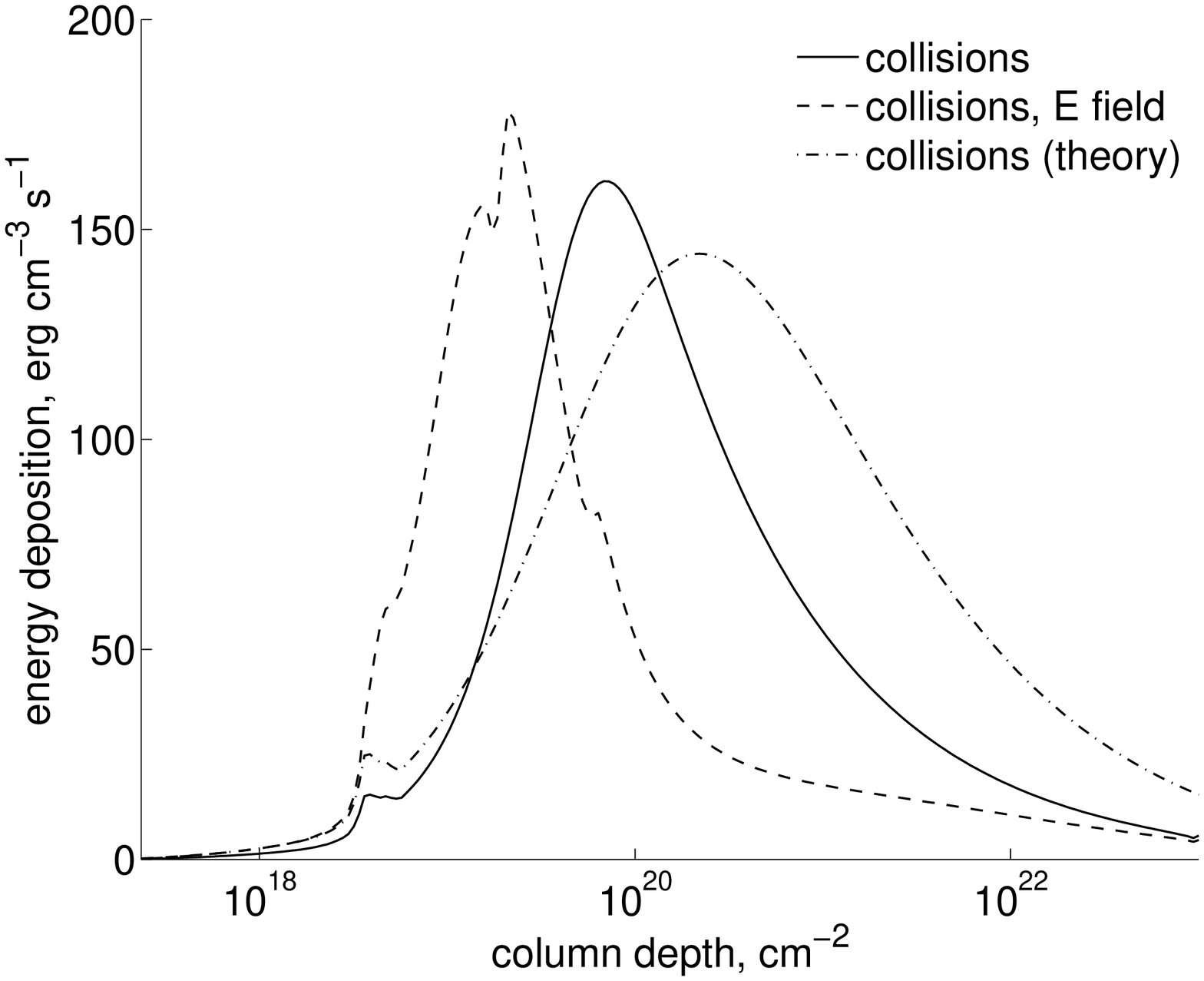}}
    \label{fig:heat-const-all}}
  \caption{Pitch angle distribution (a) and heating function of the beam (b). Beam parameters are the same as in Fig.~\ref{fig:relax} and the magnetic convergence (for dotted curve) is given by Eq.~(\ref{eq:conv3}).}
\end{figure}

The number of electrons returning back to the source plotted in Fig.~\ref{fig:mu-const-all} is smaller compared to the case when the self-induced electric field is not taken into account. However, even without the electric field $\varepsilon$, a number of electrons with $\mu<0$ is essential owing to the pitch angle scattering (second and third terms on the right hand side of Eq.~(\ref{eq:fok_pl})).

The heating function plot (Fig.~\ref{fig:heat-const-all}) shows that if a self-induced electric field is taken into account the heating is most effective at the upper column depth ($2 \tento{19} \percmsq$) compared to a much lower one ($10^{20} \percmsq$) in the pure collisional beam relaxation, which is consistent with the results obtained by \citet{Emslie80}. Indeed, the inclusion of the electric field decreases the stopping depth \citep{Zharkova06} and increases the number of returning electrons, thus, reducing the number of electrons at larger depths. All these factors lead to the upward shift of the heating function maximum. The theoretical heating curve for pure collisions \citep{Syrovatskii72} is plotted in Fig.~\ref{fig:heat-const-all}.

\subsection{Effects of a magnetic convergence}

\begin{figure}
  \centering
  \resizebox{\hsize}{!}{\includegraphics{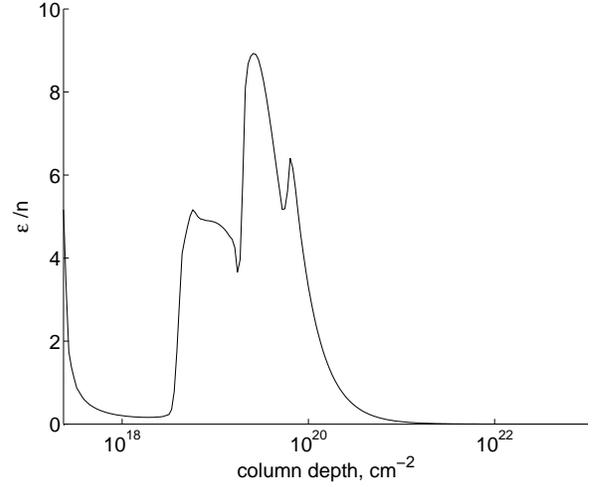}}
  \caption{The ratio of the dimensionless electric field to the ambient plasma density as a function of column depth. Collisions and electric field are taken into account. Beam parameters are the same as in Fig.~\ref{fig:relax}.}
  \label{fig:E-n-const-Bc0}
\end{figure}

\begin{figure}
  \centering
  \subfloat[]{
    \resizebox{\hsize}{!}{\includegraphics[width=0.45\textwidth]{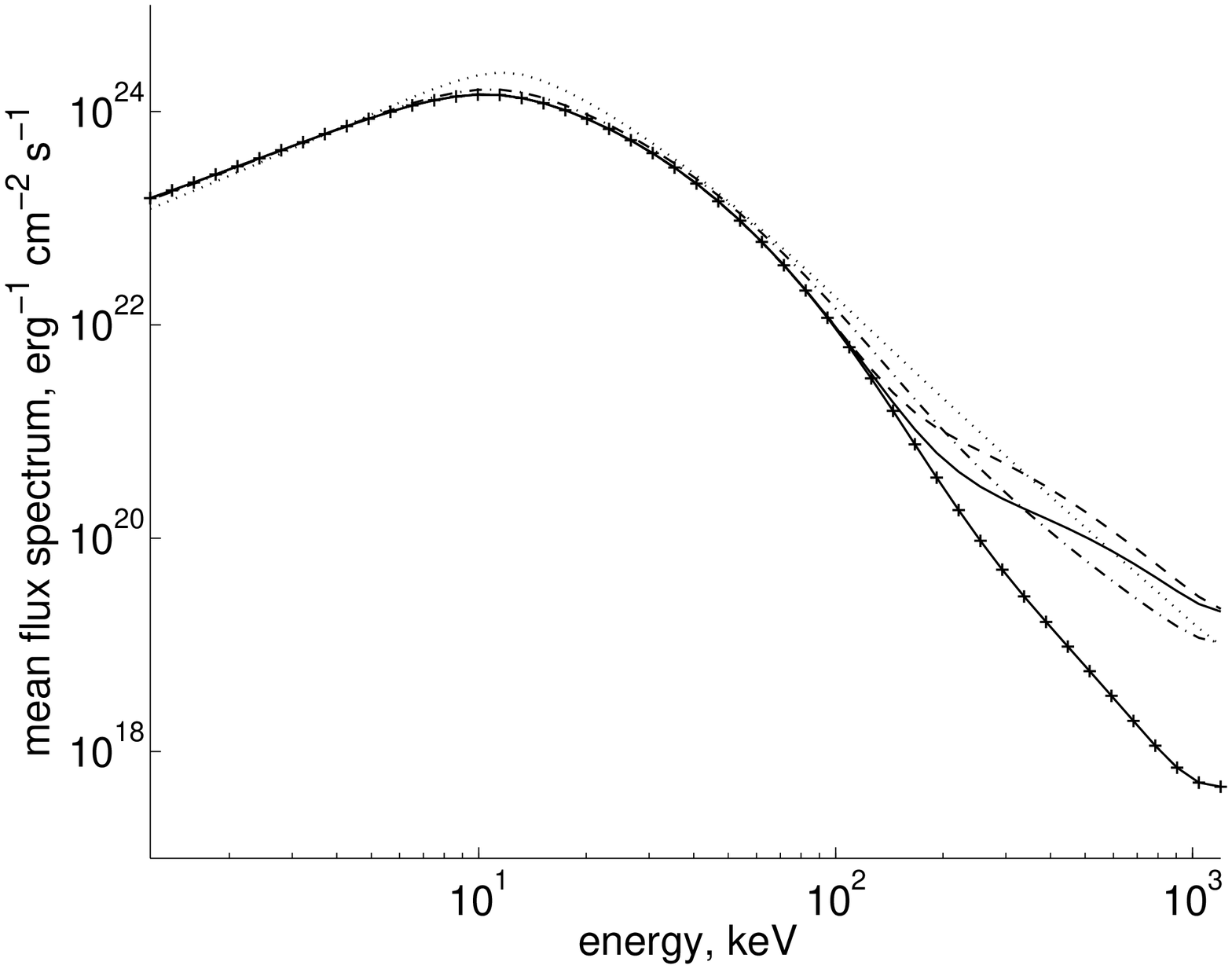}}
    \label{fig:sp-const-conv}}\quad
  \subfloat[]{
    \resizebox{\hsize}{!}{\includegraphics[width=0.45\textwidth]{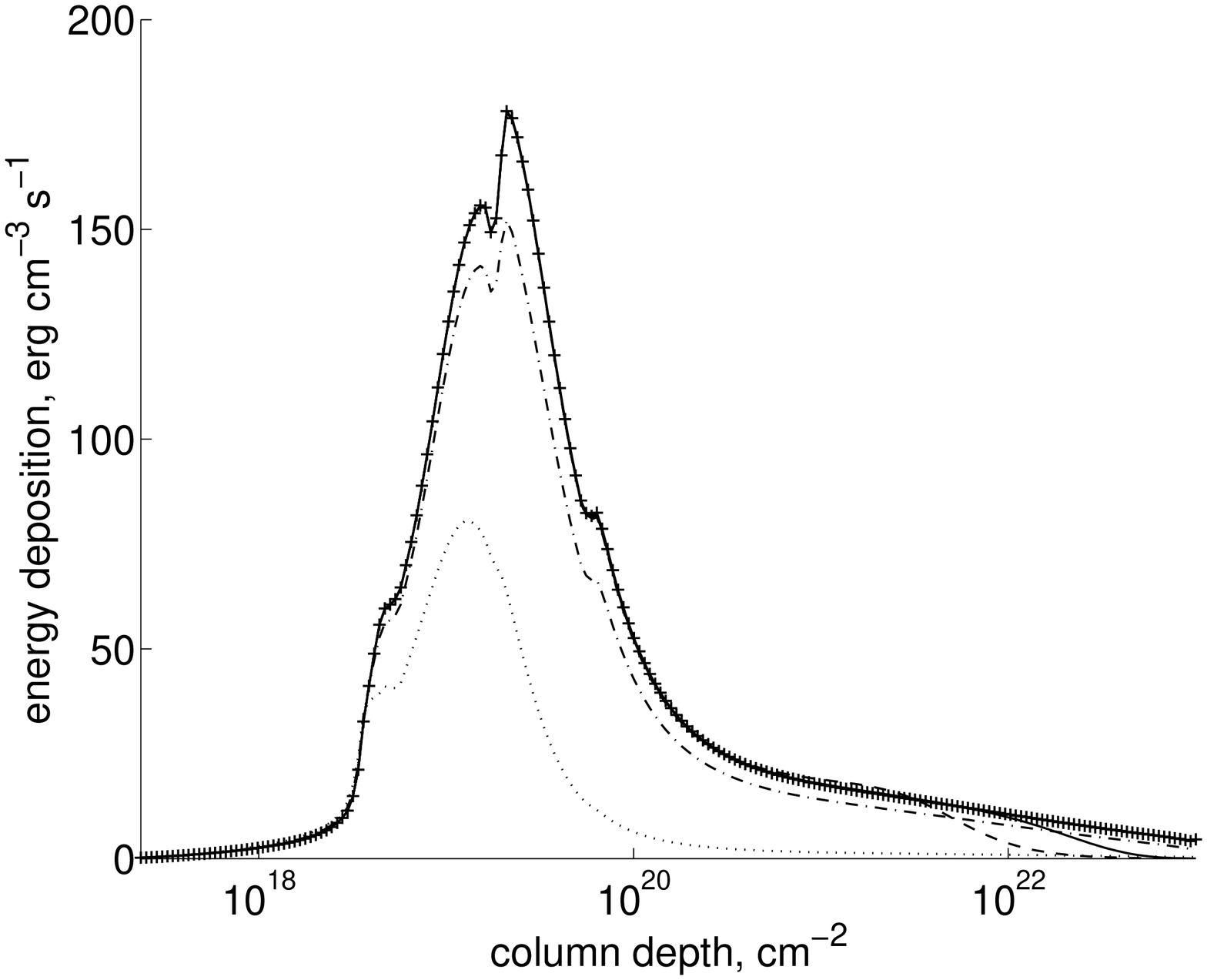}}
    \label{fig:heat-const-conv}}
  \caption{Mean flux spectra (a) of the upward ($\mu<0$) propagating electrons and energy deposition (b) without (crosses) and with magnetic convergence given by Eq.~(\ref{eq:convB1}) (solid curve), Eq.~(\ref{eq:convB2}) (dashed curve), Eq.~(\ref{eq:convB3}) (dotted curve) and Eq.~(\ref{eq:conv4}) (dot-dashed curve). Beam parameters are the same as in Fig.~\ref{fig:relax}.}
  \label{fig:all-const-conv}
\end{figure}

The converging magnetic field acts as a magnetic mirror and can essentially increase the number of the electrons that move upwards. Let us determine how large the magnetic convergence parameter, $\alpha_\mathrm{B}$, should be to have any noticeable effect on the distribution of beam electrons. In order to do so we compare the terms in front of $\partial f/\partial\mu$ in Eq.~(\ref{eq:fok_pl1}). Since the collisional pitch angle diffusion is much smaller than the one caused by the electric field (see Fig.~\ref{fig:mu-const-all}), we compare the effects of the magnetic convergence with those caused by the electric field. The magnetic convergence effects are stronger if $\alpha_\mathrm{B} > 2 \varepsilon /(n z)$. To estimate this expression we plot the dimensionless ratio $\varepsilon /n$ in Fig.~\ref{fig:E-n-const-Bc0}. The minimal value of the ratio $\varepsilon /n$ in the interval from $s=s_\mathrm{min}$ to the stopping depth of low energy ($z=1$) electrons, $\sim 5 \tento{19} \percmsq$, is about $0.2$. Thus, the magnetic convergence would be more effective than the electric field for the electrons with energies higher than the cut-off energy if $\alpha_\mathrm{B} \gtrsim 0.4$. High energy electrons can travel much deeper into the chromosphere (see Fig.~\ref{fig:loss-const-Bc0-noE}), where the ratio $\varepsilon /n$ can be as low as $2 \tento{-4}$, thus the magnetic convergence would be more effective for them if $\alpha_\mathrm{B} \gtrsim 4\tento{-6}$.

In the following subsections we present the results of simulations for different models of the converging magnetic field. These results are illustrated by the mean flux spectra plots (Fig.~\ref{fig:sp-const-conv}) for the upward ($\mu<0$) moving electrons, while the spectra of the downward moving electrons are found to be very close for all convergence models. The energy deposition profiles for different magnetic field approximations are shown in Fig.~\ref{fig:heat-const-conv}.

\subsubsection{Exponential approximation of a magnetic field convergence}

Following the approximation proposed by \citet{Leach81}, let us assume that the convergence parameter does not depend on depth
\begin{equation} \label{eq:conv1}
    \alpha_\mathrm{B} = \alpha_{B0} = const,
\end{equation}
then the magnetic field variation is
\begin{equation} \label{eq:convB1}
    B(s) = B_0 \exp \left(\alpha_{B0} \left(s-s_\mathrm{min}\right) \right).
\end{equation}

Suppose that the magnetic field at the depth $s_\mathrm{max}$ is 1000 times stronger than at the depth $s_\mathrm{min}$, then $\alpha_{B} \approx \ln(1000)/s_\mathrm{max} = 7.5 \tento{-3}$. As we discussed earlier the effect of the magnetic convergence with such low $\alpha_{B}$ would be noticeable only for high energy ($> 100 \units{keV}$) electrons. This is clearly illustrated by a comparison of the mean flux spectra of the moving upward electrons and heating function obtained with and without magnetic convergence (see solid and dashed curves in Fig.~\ref{fig:all-const-conv}). While the electric field returns mostly the low and mid energy electrons and makes the spectrum of the returning electrons softer (in comparison with the purely collisional case), the magnetic mirror returns back the high energy electrons and makes their spectrum harder and similar to the initial power law. On the other hand, magnetic convergence reduces the heating at the deeper layers (Fig.~\ref{fig:heat-const-conv}), where it is caused by high energy electrons, because they were mirrored back to the corona.

\subsubsection{Parabolic approximation}

\citet{McClements92} suggested the following profile of a magnetic field variation:
\begin{equation}\label{eq:convB2}
    B(s) = B_0 \left( 1 + \frac{(s-s_\mathrm{min})^2}{s_0^2} \right).
\end{equation}
If $B(s_\mathrm{max})/B(s_\mathrm{min}) = 1000$ then $s_0 \approx s_\mathrm{max}/\sqrt(1000) = 31.6$. The convergence parameter is
\begin{equation} \label{eq:conv2}
    \alpha_\mathrm{B}(s) = \frac{2(s-s_\mathrm{min})}{s_0^2+(s-s_\mathrm{min})^2}.
\end{equation}
The magnetic convergence parameter $\alpha_{B}$ at maximum is $1/s_0 \approx 3.16 \tento{-2}$. This value is not high enough to affect all the beam electrons, but as is seen in Fig.~\ref{fig:all-const-conv} this convergence model, in a similar way as the previous one, increases the number of high energy electrons with negative $\mu$ and reduces the heating of deep atmosphere layers. Note, also, that \cite{McClements92} considered a constant plasma density, and in their model the profile of the magnetic field given by Eq.~(\ref{eq:convB2}) can be more effective, while in our case plasma density exponentially increases with depth. Here we assumed that the magnetic field changes according to Eq.~(\ref{eq:convB2}) in the whole range of column depths. However it would be more appropriate to assume that the magnetic field variation is different in the corona and chromosphere. Such an approach is discussed in the following section.

\subsubsection{Hybrid approximation of magnetic field} \label{sec:var_b}

\begin{figure}
  \centering
  \subfloat[convergence]{
    \resizebox{\hsize}{!}{\includegraphics{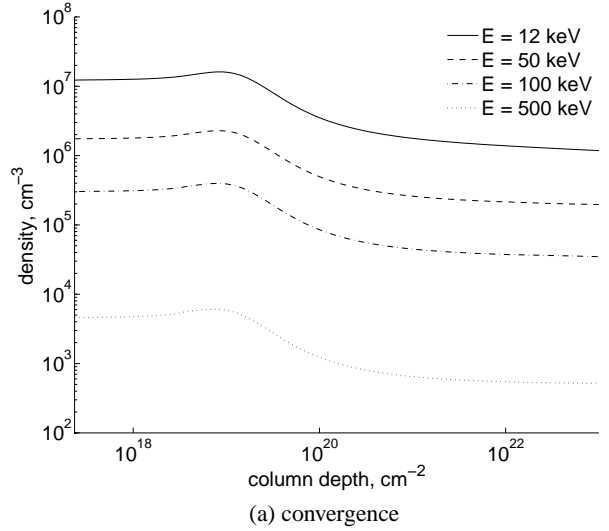}}
    \label{fig:loss-const-Bc10-noE-noC}}\quad
  \subfloat[convergence and collisions]{
    \resizebox{\hsize}{!}{\includegraphics{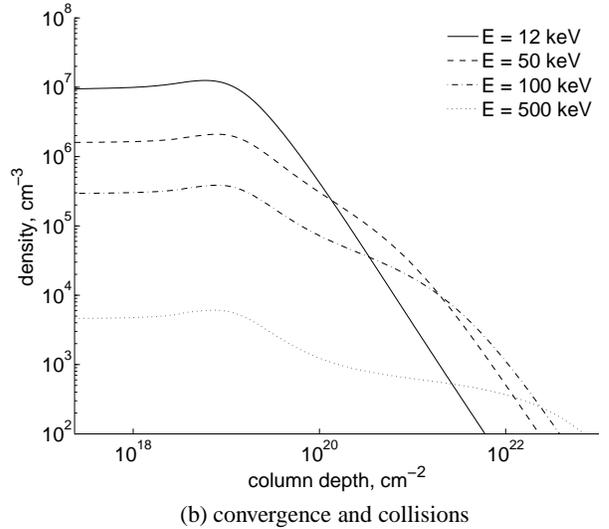}}
    \label{fig:loss-const-Bc10-noE}}
  \caption{Beam electron density in different energy bands. The magnetic convergence parameter is given by Eq.~(\ref{eq:conv3}). Beam parameters are the same as in Fig.~\ref{fig:relax}.}
  \label{fig:loss-const-Bc10-all}
\end{figure}

\begin{figure}
  \centering
  \subfloat[]{
    \resizebox{\hsize}{!}{\includegraphics{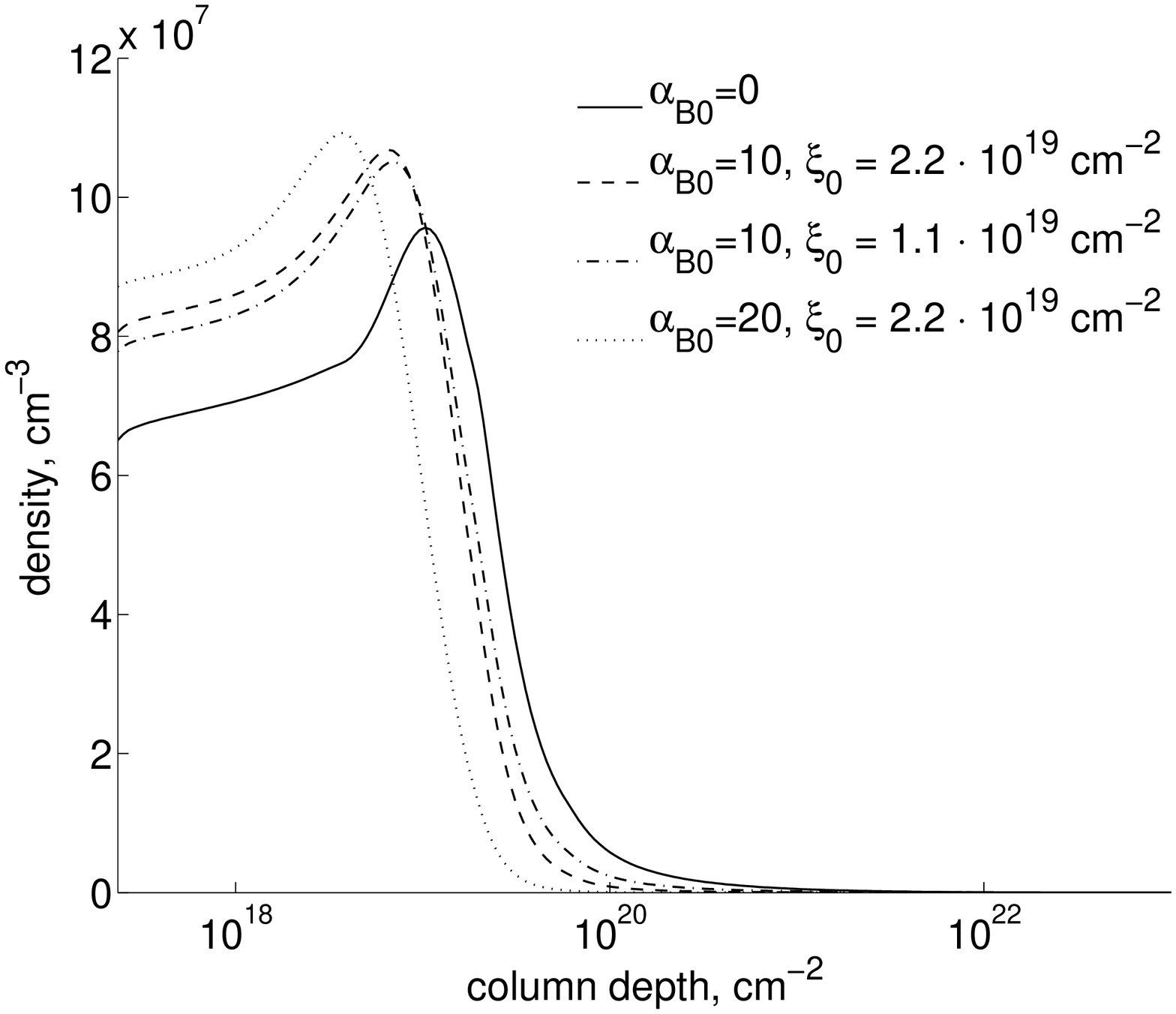}}
    \label{fig:d-const-conv1}}\quad
  \subfloat[]{
    \resizebox{\hsize}{!}{\includegraphics{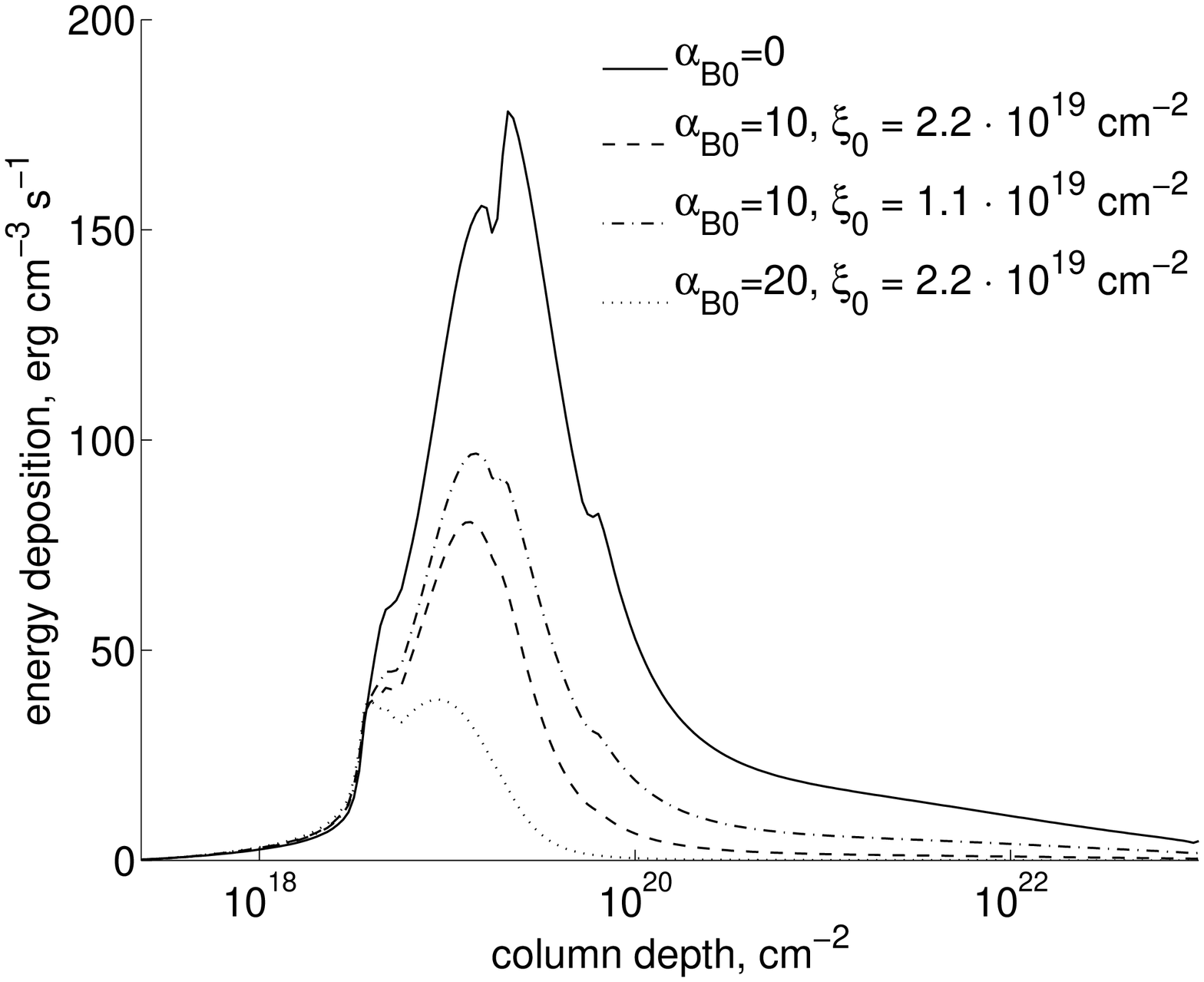}}
    \label{fig:heat-const-conv1}}
  \caption{Beam density (a) and energy deposition (b) as a function of column depth. The magnetic convergence parameter is given by Eq.~(\ref{eq:conv3}) with various $\alpha_{B0}$ and $s_0$. Beam parameters are the same as in Fig.~\ref{fig:relax}.}
  \label{fig:all-const-conv1}
\end{figure}

In this model we propose that $\alpha_\mathrm{B}$ is close to constant at small depth (in the corona) and tends to zero after some depth $s_0$ (in the chromosphere):
\begin{equation} \label{eq:conv3}
    \alpha_\mathrm{B} (s) = \alpha_{B0} \frac{s_0^2}{s_0^2+(s-s_\mathrm{min})^2},
\end{equation}
then the magnetic field variation is
\begin{equation} \label{eq:convB3}
\begin{split}
    B(s) = &B_0 \exp \left( \int\limits_{s_\mathrm{min}}^{s} \alpha\left(s^{\prime}\right) ds^{\prime} \right) = \\
           &B_0 \exp \left( \alpha_{B0} s_0 \arctan \left( \frac{s-s_\mathrm{min}}{s_0} \right) \right).
\end{split}
\end{equation}
At shallow depths, where $s \ll s_0$, the magnetic field varies as $B \approx B_0 \exp (\alpha_{B0}(s-s_\mathrm{min}))$, and at deeper layers, $s \gg s_0$, the magnetic field is constant, $B \approx B_0 \exp (\alpha_{B0} s_0 \pi/2)$. In most of the simulations (where it is not stated explicitly) we accept $\alpha_{B0}=10$ and $s_0=0.2$ (or $2.2\tento{19}\percmsq$, which corresponds to the transitional region), this makes the ratio $B(s_\mathrm{max})/B(s_\mathrm{min})$ to be equal $23.1$.

Electrons with velocities inside the loss-cone are not reflected by the magnetic mirror and reach the deep layers. For the current convergence model the critical pitch angle cosine of the loss-cone is $\mu_\mathrm{lc} = \sqrt{1-B(s_\mathrm{min})/B(s_\mathrm{max})} \approx 0.98$. This means that for the accepted initial angle dispersion of $0.2$ about $90\%$ of the electrons are reflected back.

As it is seen in Fig.~\ref{fig:heat-const-all} the effect of magnetic convergence on the pitch angle distribution is similar to the effect of the electric field. However, since the electric field is most effective for the electrons with $\mu=\pm 1$, the pitch angle distribution has a maximum at $\mu = 1$ when the convergence is not taken into account. On the contrary, the magnetic field does not affect electrons moving along the field lines, thus, the angle distribution of the upward moving electrons has a maximum at $\mu_\mathrm{m} \approx -0.8$ (Fig.~\ref{fig:mu-const-all}), which is consistent with the conclusions of \citet{Zharkova06}.

As the convergence parameter $\alpha_\mathrm{B}$ is relatively high in this model, the whole spectrum of electron energy is affected (see Fig.~\ref{fig:sp-const-conv}). The energy deposition profile for this magnetic field approximation (Fig.~\ref{fig:heat-const-conv}) indicates that the heating is only about $30\%$ of the heating produced in the case of constant magnetic field, which is because many of electrons are reflected by the magnetic mirror before they get into dense plasma.

The profiles of electron density with different energies are plotted in Fig.~\ref{fig:loss-const-Bc10-all}. If only magnetic convergence is taken into account (Fig.~\ref{fig:loss-const-Bc10-noE-noC}), it can be seen that magnetic mirroring does not depend on the electron energy. Electrons the pitch angles of which are outside the loss-cone are turned back at depth $\sim 10^{19} - 10^{20} \percmsq$. The remaining electrons (the pitch angles of which are inside the loss-cone) can travel down to the lower boundary in the atmosphere. When the collisions are taken into account, electrons, especially those with low energies, lose their energy due to collisions (Fig.~\ref{fig:loss-const-Bc10-noE}). It is important to compare collisional beam relaxation with and without magnetic convergence plotted in Figs.~\ref{fig:loss-const-Bc10-noE} and \ref{fig:loss-const-Bc0-noE} respectively. It is noticeable that the combination of the effects of collisions and convergence is stronger than the sum of the two separate effects. This occurs because electrons with the initial pitch angles inside the loss-cone are scattered by collisions to pitch angles which fall out from the loss-cone, therefore, more electrons are returned back by the magnetic mirror.

In Fig.~\ref{fig:all-const-conv1} the results of simulations are presented for different magnitudes for the parameters $\alpha_{B0}$ and $s_0$ of the magnetic convergence model given by Eq.~(\ref{eq:conv3}). The increase of $\alpha_{B0}$ clearly affects the beam electrons by reducing the depth of their penetration and by increasing the number of returning electrons. It is obvious that the system would be sensitive to the variation of $s_0$ if it is smaller than the penetration (stopping) depth. This is proven in Fig.~\ref{fig:d-const-conv1}.

\subsubsection{Magnetic field model fitted to the observations}

Although the magnetic field cannot be directly measured in the solar atmosphere, there are some indirect techniques which allow the magnitude of the magnetic field to be estimated. The coronal magnetic field can be determined from radio observations of gyro-resonance emission \citep{Lang93, Brosius06}. In particular, \citet{Brosius06} suggest that the magnetic scale height above sunspots, $L_{B\mathrm{cor}} = B/\Delta B$, is $\sim 7 \units{Mm}$. On the other hand, \citet{Kontar08} determined the chromospheric magnetic field by measuring the sizes and heights of hard X-ray sources in different energy bands. They found that the chromospheric magnetic scale height, $L_{B\mathrm{chr}}$, is $\sim 0.3 \units{Mm}$. Assuming that the magnetic scale height, $L_B$, changes linearly with depth $l$, the convergence parameter is
\begin{equation} \label{eq:conv4}
    \alpha_\mathrm{B} (s) = \frac{\alpha_{B0}}{n L_B},
\end{equation}
where $\alpha_{B0} = E_0^2 / (\pi e^4 n_0 \ln\Lambda L_{B\mathrm{cor}}) = 15.7$, and the dimensionless magnetic scale height as a function of the column depth is
\begin{equation}
    L_B = 1 - \left( 1 - \frac{L_{B\mathrm{chr}}}{L_{B\mathrm{cor}}}\right) \frac{l(s)}{l(s_\mathrm{max})},
\end{equation}
where the linear depth is $l(s) \propto \int n(s)^{-1} ds$. Since the magnetic field is defined as function of the linear depth the model depends on the density profile $n(s)$ of the background plasma.

As it is seen in Fig.~\ref{fig:sp-const-conv}, such the magnetic convergence affects the electrons with all energies. This leads to about $20\%$ reduction of the heating produced by beam electrons in comparison with constant magnetic field profile (Fig.~\ref{fig:heat-const-conv}).

\section{Impulsive injection} \label{sec:impulse}

\begin{figure}
  \centering
  \subfloat[]{
    \resizebox{\hsize}{!}{\includegraphics{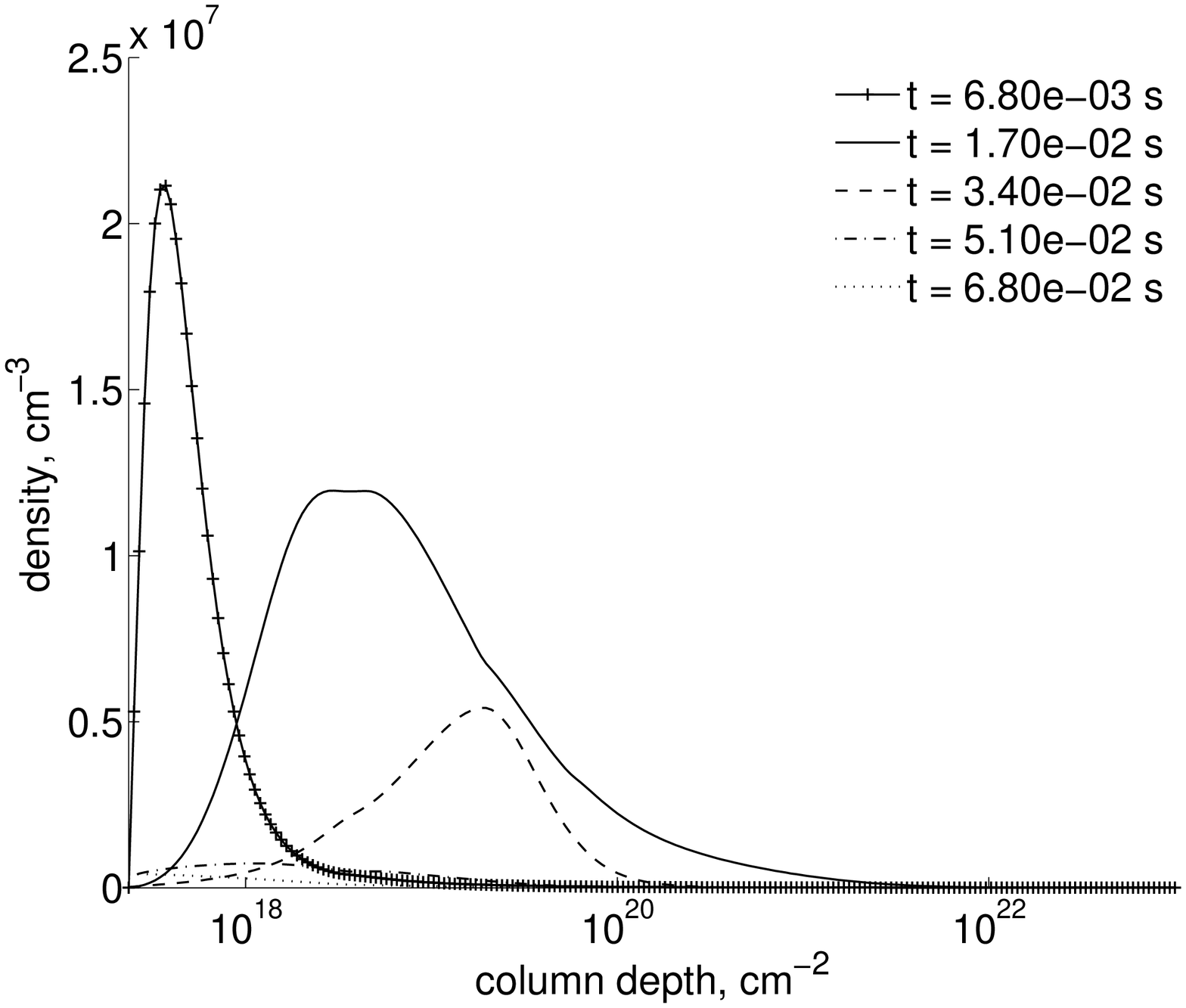}}
    \label{fig:d-puls10-Bc0}} \quad
  \subfloat[]{
    \resizebox{\hsize}{!}{\includegraphics{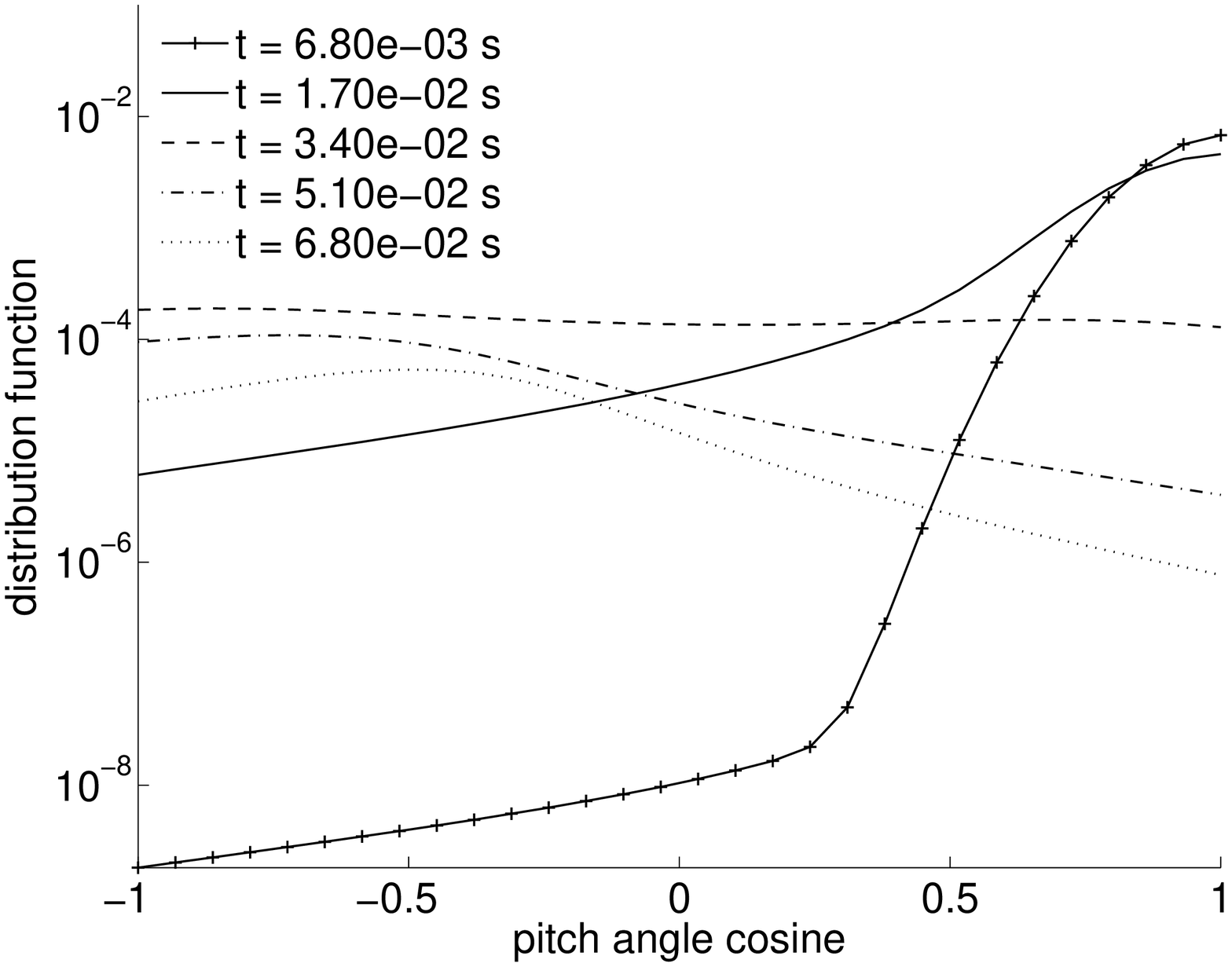}}
    \label{fig:mu-puls10-Bc0}} \\
  \caption{Electron density at impulsive injection as a function of column depth (a) and pitch angle (b). Collisions and the electric field are taken into account. The beam parameters, see Eq.~(\ref{eq:initdistr}), are $\gamma=3$, $F_\mathrm{top}=10^{10} \fluxunit$, $\Delta\mu =0.2$ and injection time, $\delta t = 1.7\tento{-3}\units{s}$.}
  \label{fig:all-puls10-Bc0}
\end{figure}

\begin{figure*}
  \centering
  \subfloat[collisions]{
    \includegraphics[width=0.45\textwidth]{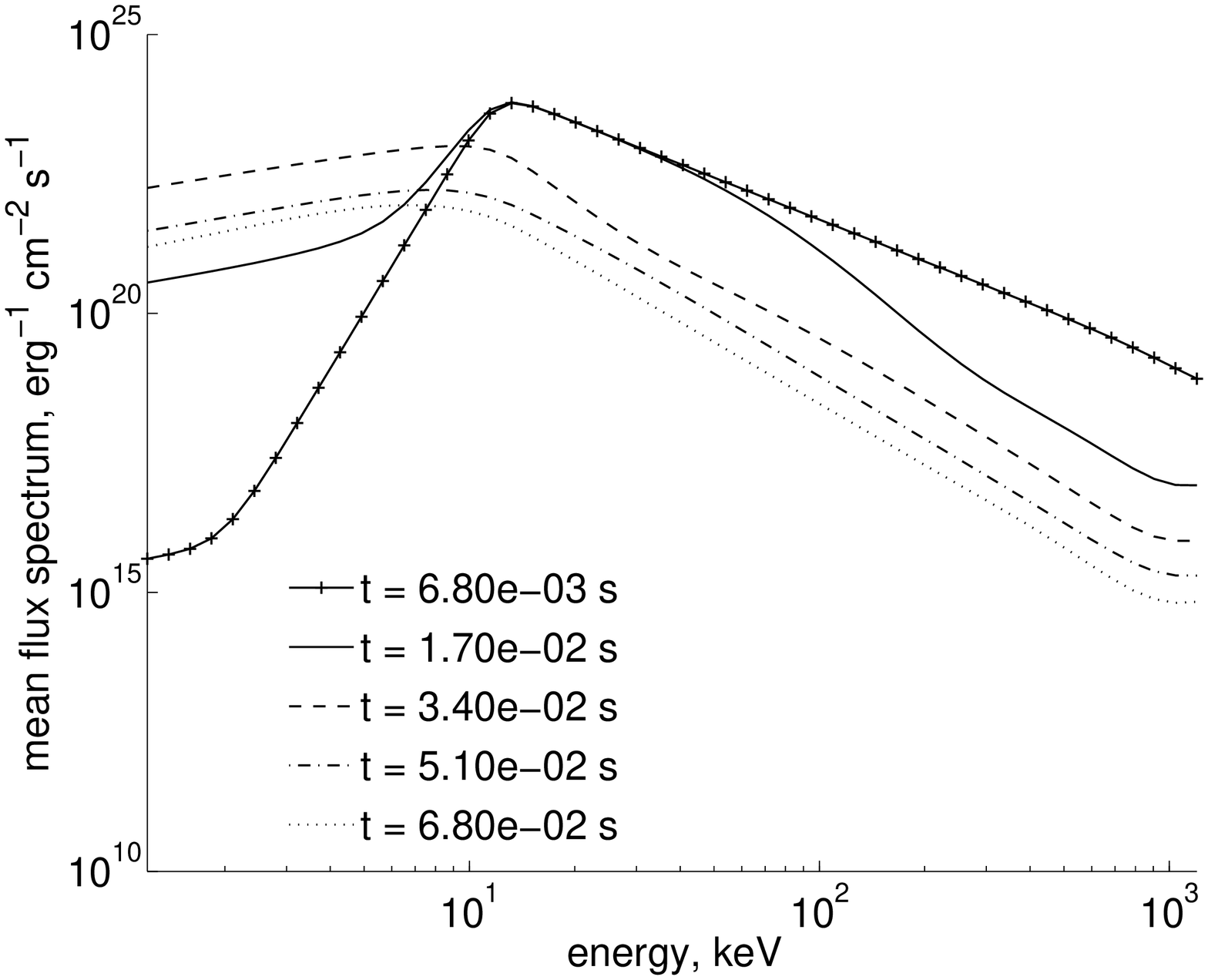}
    \label{fig:sp-puls10-Bc0-noE}}\quad
  \subfloat[collisions and electric field]{
    \includegraphics[width=0.45\textwidth]{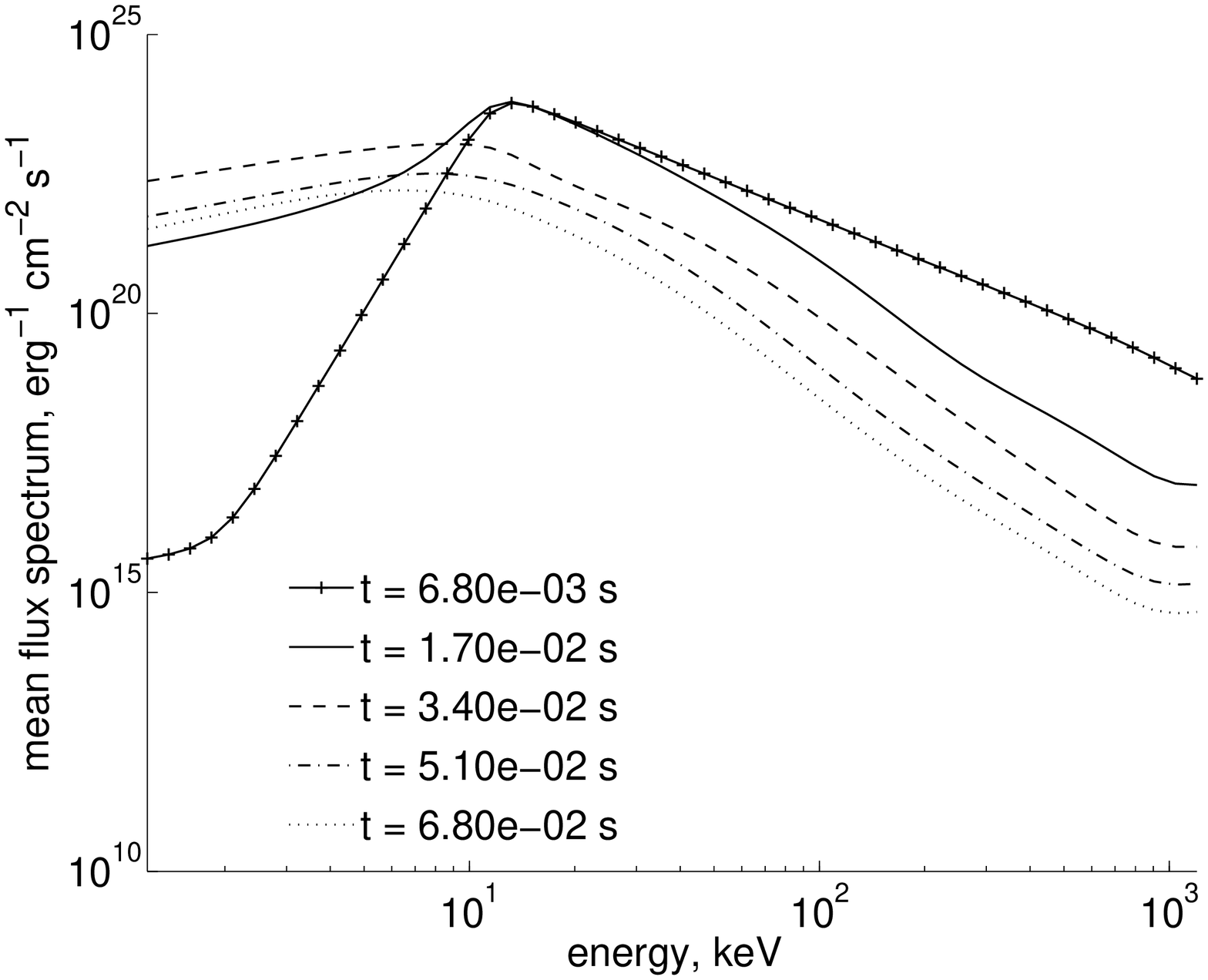}
    \label{fig:sp-puls10-Bc0}} \\
  \subfloat[collisions and convergence according to Eq.~(\ref{eq:conv3})]{
    \includegraphics[width=0.45\textwidth]{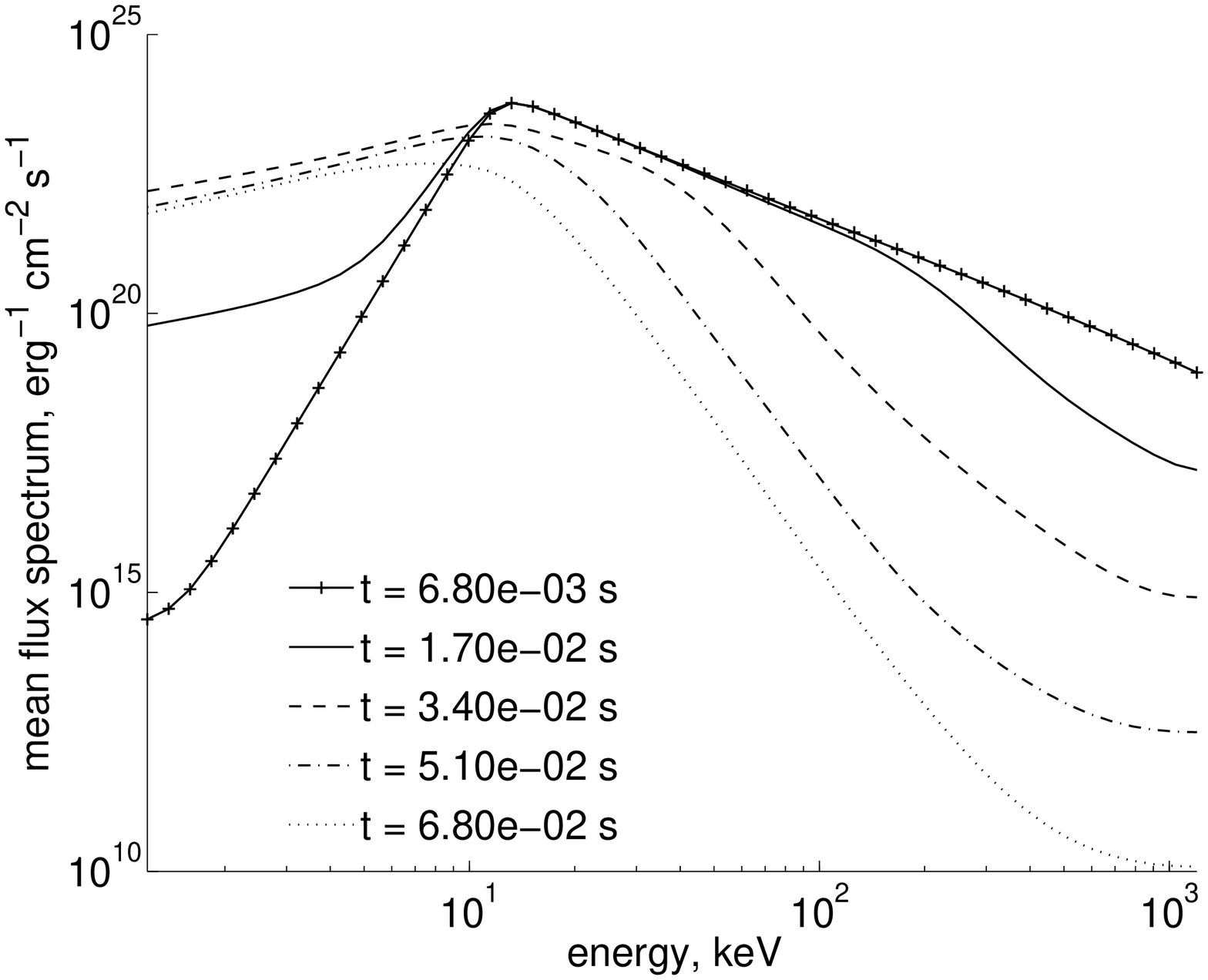}
    \label{fig:sp-puls10-Bc10-noE}}
  \subfloat[collisions and convergence according to Eq.~(\ref{eq:conv4})]{
    \includegraphics[width=0.45\textwidth]{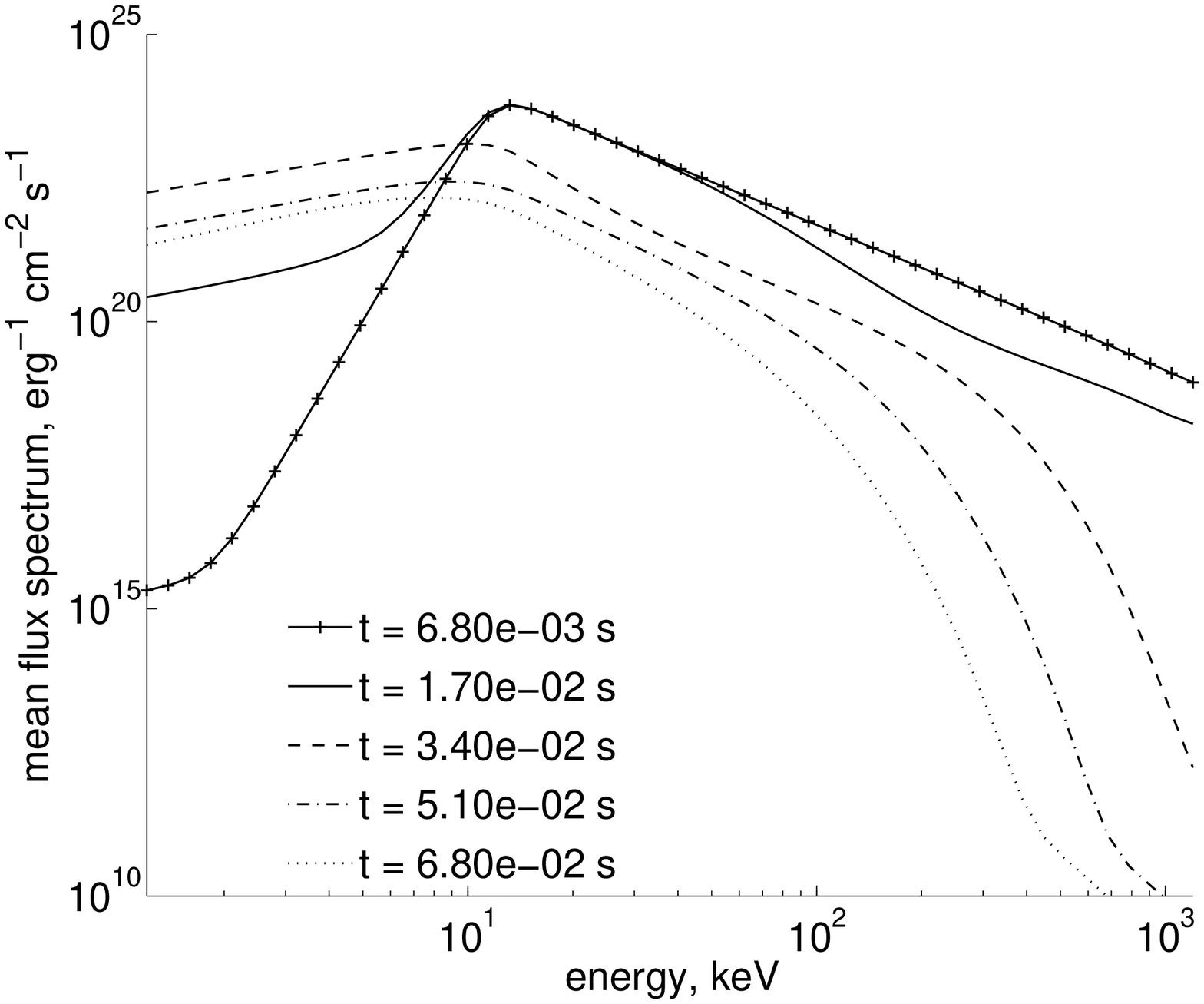}
    \label{fig:sp-puls10-Bcq-noE}}
  \caption{Mean flux spectra of the electrons injected as a short impulse. The beam parameters are the same as in Fig.~\ref{fig:all-puls10-Bc0}.}
  \label{fig:sp-puls10-all}
\end{figure*}

Now we can consider the precipitation of electron beam injected as a short impulse. The early SMM \citep{Kiplinger83} and recent CORONAS/IRIS \citep{Charikov04} observations reveal millisecond impulses in the hard X-ray emission from solar flares. Furthermore, the electron acceleration time in a reconnecting current sheet was recently shown \citet{Siversky09} to be as short as $10^{-5} \units{s}$. These facts suggest that the time scale of a beam of accelerated electrons may be rather short. In this section we study the evolution of such short impulse in the solar atmosphere. The injection time, $\delta t$, is chosen to be $1.7\tento{-3}\units{s}$, which is much shorter than the relaxation time $t_\mathrm{r} \approx 0.07 \units{s}$, that was found in Sec.~\ref{sec:relax} (see Fig.~\ref{fig:relax}). The default parameters of the beam are similar to the case of the stationary injection: the initial spectral index of the beam is $\gamma = 3$, the maximal energy flux at the top boundary is $F_\mathrm{top} = 10^{10} \fluxunit$ and the initial angle dispersion is $\Delta\mu = 0.2$. Also the energy deposition profiles produced both by a softer beam ($\gamma = 7$) and by a stronger beam ($F_\mathrm{top} = 10^{12} \fluxunit$) are obtained.

The impulse injection, obviously, leads to a smaller density of electrons at a given depth in comparison with the stationary injection (see Figs.~\ref{fig:relax_f} and \ref{fig:d-puls10-Bc0}). A smaller density results in a lower self-induced electric field. Thus, in the case of a short impulsive injection the electric field does not affect so much the distributions. As a result, the only mechanism that can essentially increase the number of returning electrons is a magnetic convergence. As it was mentioned above in the current study we assume that the beam current is always compensated by the plasma return current, thus the self-induced electric field develops immediately. However, as was shown by \citet{Oord90}, the neutralisation time of the beam current is of the order of the collisional time. Thus, the effect of electric field for short impulses can be even smaller than our estimations.

Anisotropic scattering of beam electrons in collisions with the ambient plasma makes the pitch angle distribution more flat with time (see Fig.~\ref{fig:mu-puls10-Bc0}). The electrons propagating downward reach depths with high density of the ambient plasma, lose their energy due to collisions and leave the distribution (become thermalised) when their energy is less than $z_\mathrm{min}$. On the contrary, the returning electrons move into less dense plasma almost without losing any energy, but gaining it in the self-induced electric field. Thus, after some time the number of upward moving electrons can exceed the number of downward moving ones, which is clearly seen in Fig.~\ref{fig:mu-puls10-Bc0}. The angle distributions show that after $\sim 3.4 \tento{-2} \units{s}$ most of the downward propagating electrons are gone and the majority of electrons have $\mu<0$, i.e. they move back to the source in the corona.

\subsection{Energy spectra}

Since the first term at the right hand side of Eq.~(\ref{eq:fok_pl}), which is responsible for the energy losses due to collisions, is proportional to $z^{-1/2}$ (where $z$ is the dimensionless energy), one might expect that electron spectra would become harder with time. However, the downward moving electrons with higher energy reach the dense plasma faster and lose their energy faster than lower energy electrons, which makes the energy spectra softer with time (Fig.~\ref{fig:sp-puls10-Bc0-noE}). The same is valid for the spectra of the upward moving electrons. In this case, the high energy electrons escape the distribution faster by reaching the top boundary ($s=s_\mathrm{min}$). Due to this effect the power law index can increase from the initial value $3$ up to $4$ during the beam evolution (Fig.~\ref{fig:sp-puls10-Bc0-noE}).

\begin{figure*}
  \centering
  \subfloat[collisions]{
    \includegraphics[width=0.45\textwidth]{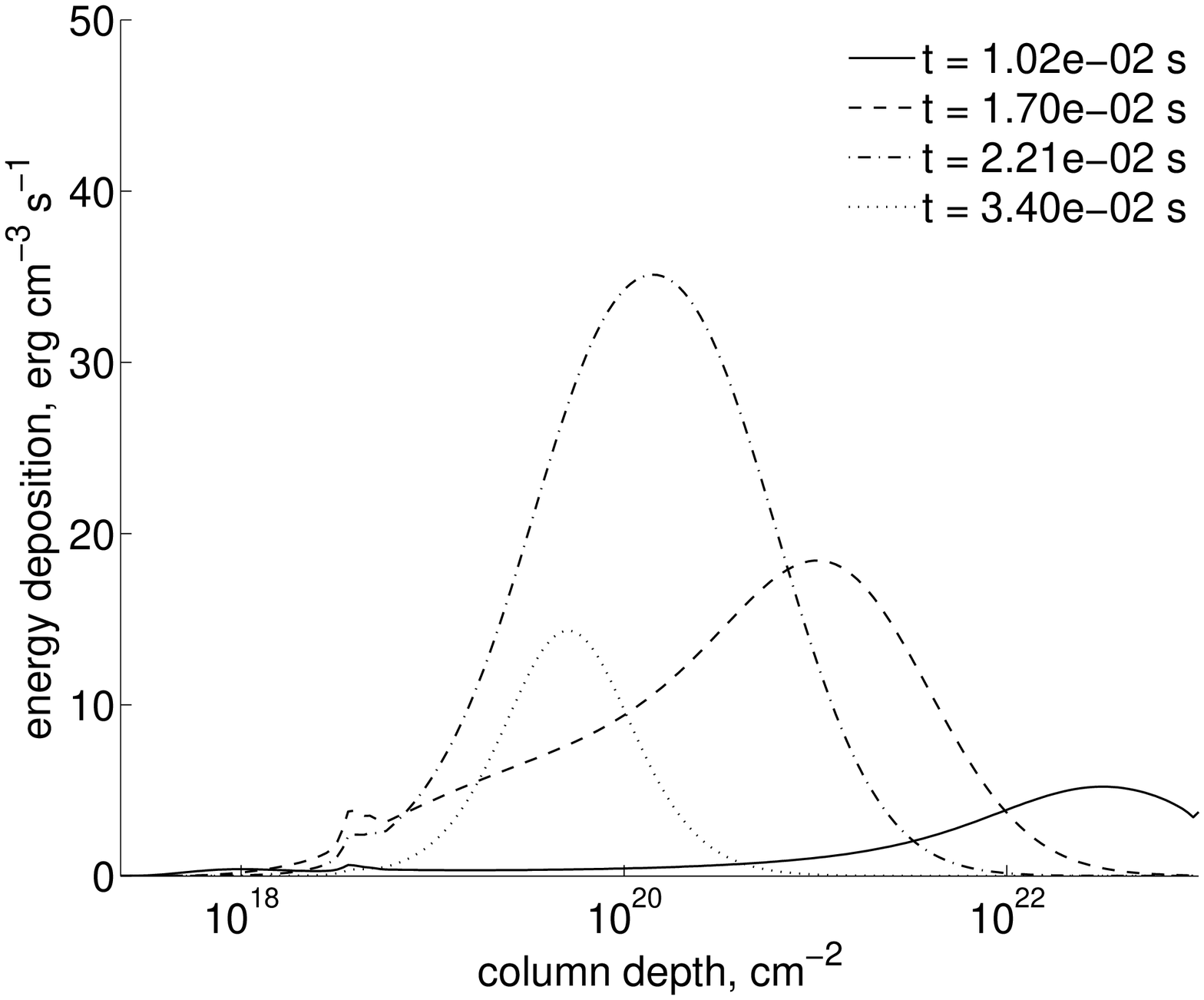}
    \label{fig:heat-puls10-Bc0-noE}} \quad
  \subfloat[collisions and electric field]{
    \includegraphics[width=0.45\textwidth]{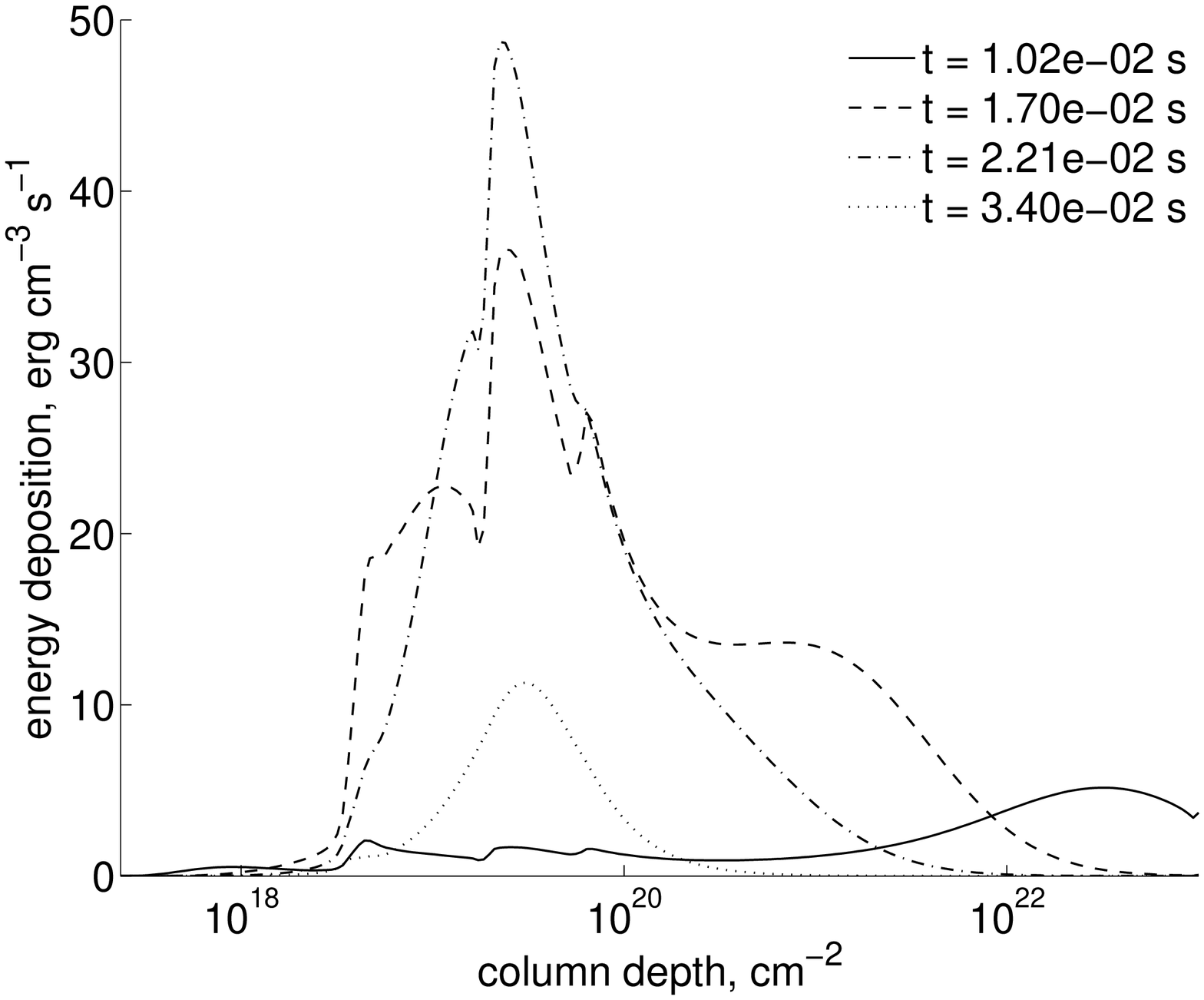}
    \label{fig:heat-puls10-Bc0}} \\
  \subfloat[collisions and convergence]{
    \includegraphics[width=0.45\textwidth]{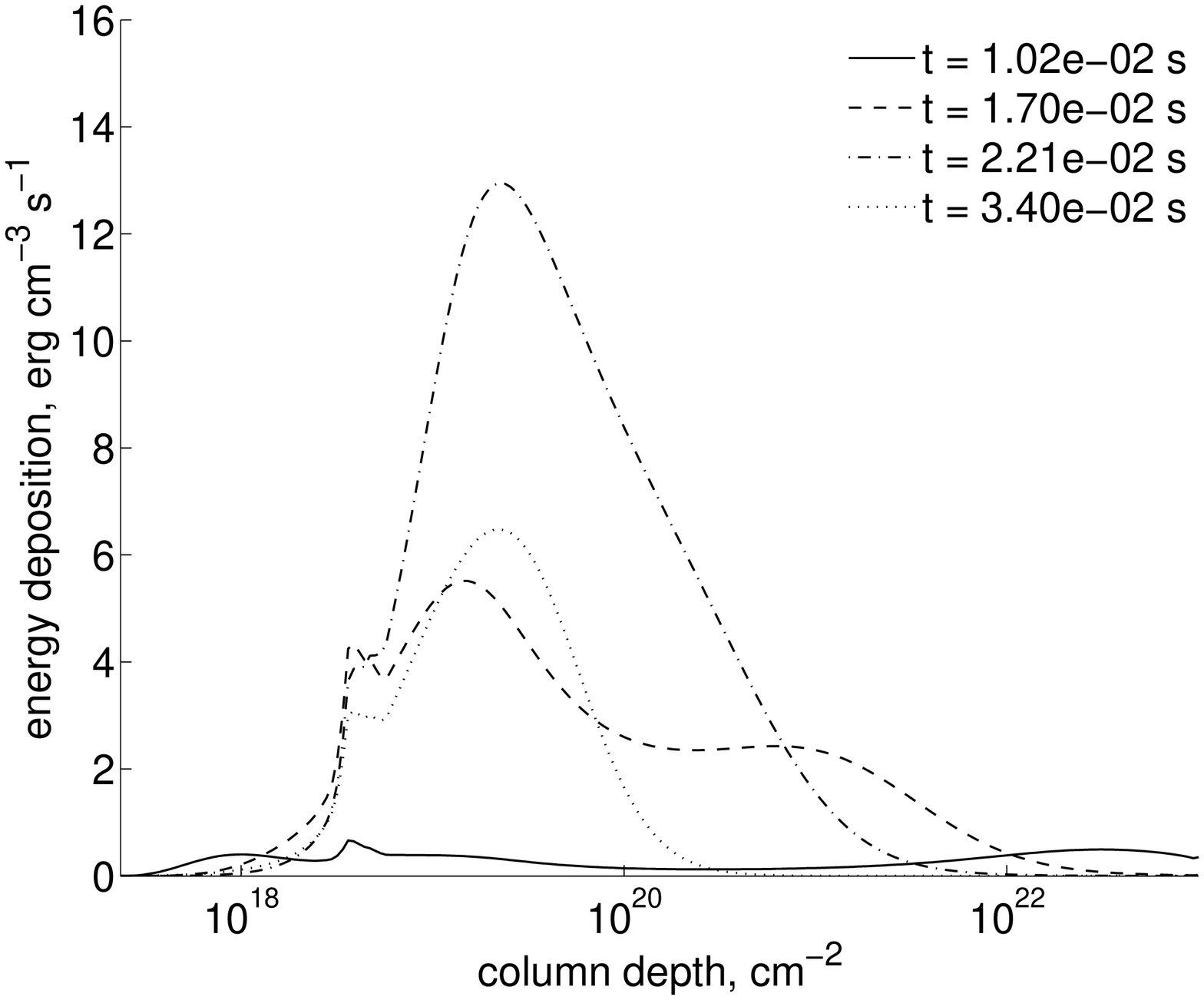}
    \label{fig:heat-puls10-Bc10-noE}} \quad
  \subfloat[collisions, convergence and electric field]{
    \includegraphics[width=0.45\textwidth]{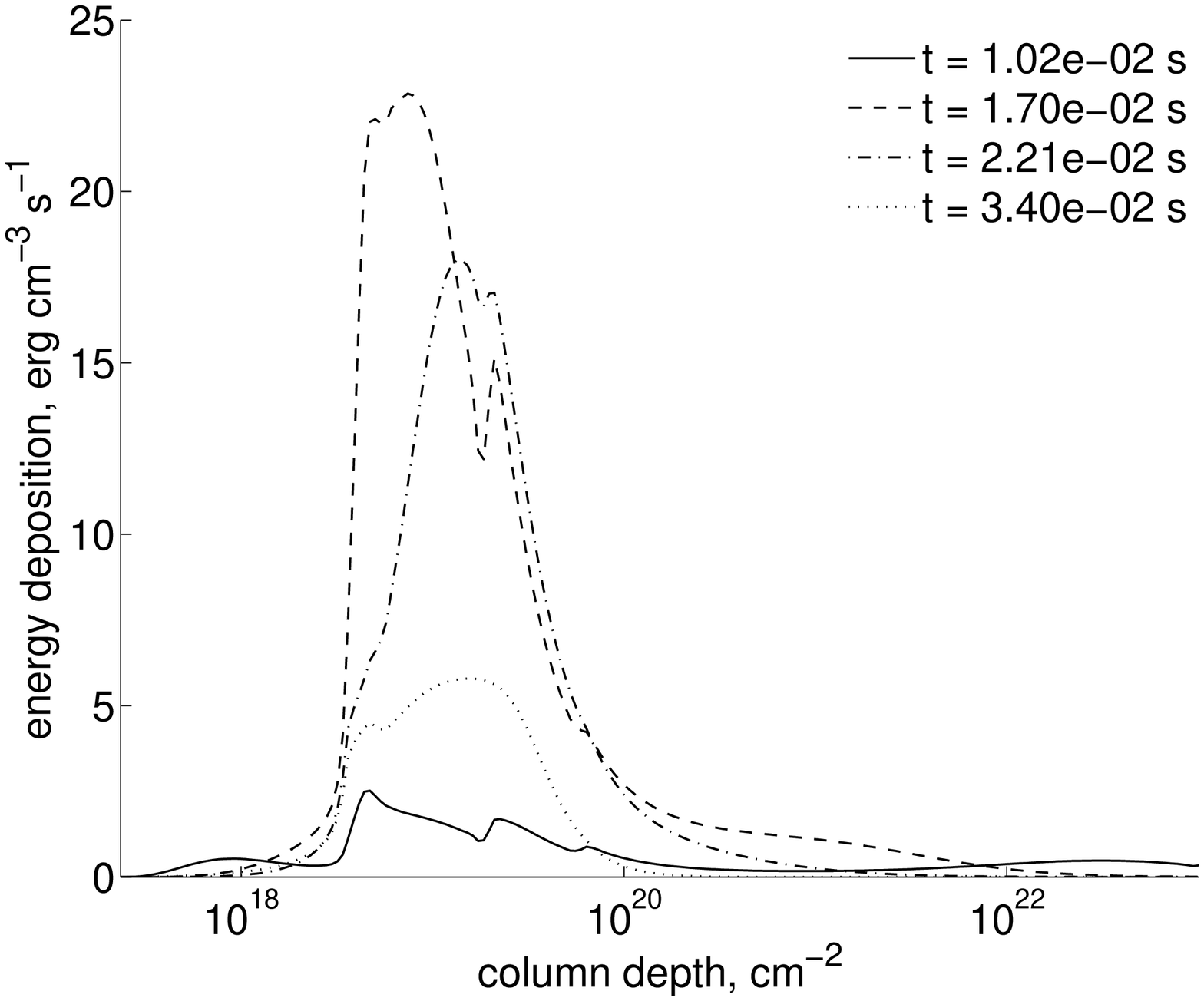}
    \label{fig:heat-puls10-Bc10}}
  \caption{Energy deposition by a beam impulse. The beam parameters are the same as in Fig.~\ref{fig:all-puls10-Bc0} and the magnetic convergence is given by Eq.~(\ref{eq:conv3}).}
  \label{fig:heat-puls10-all}
\end{figure*}

Magnetic field on its own cannot change the energy of electrons. However, a converging magnetic field acts as a magnetic mirror and returns the essential part of electrons back to the source. As was shown above in Fig.~\ref{fig:sp-const-conv} for the high energy electrons the magnetic convergence is more effective than the electric field and pitch angle diffusion. Thus, high energy electrons can quickly escape through the $s=s_\mathrm{min}$ boundary and the power law index can reach higher values than in the case with the constant magnetic field. For example, for the magnetic field profile given by Eq.~(\ref{eq:convB3}) the power law index increases from the initial value $3$ up to $8$ (Fig.~\ref{fig:sp-puls10-Bc10-noE}). If the convergence parameter is defined by Eq.~(\ref{eq:conv4}), the initial power law distribution converts to some kind of quasi-thermal distribution with an essential drop in high energies (Fig.~\ref{fig:sp-puls10-Bcq-noE}).

\subsection{Energy deposition}

Fig.~\ref{fig:heat-puls10-all} shows the evolution of energy deposition, or heating functions, when different precipitation effects are taken into account. In the purely collisional case (Fig.~\ref{fig:heat-puls10-Bc0-noE}) the heating maximum appears at the bottom boundary, moves upwards with time and vanishes near column depth $\sim 10^{19}-10^{20} \percmsq$. This evolution is consistent with stopping depths obtained for electrons with different energies (Fig.~\ref{fig:loss-const-Bc0-noE}). Indeed, the high energy electrons are the first to reach depths where the density is high enough to thermalise them. Less energetic electrons travel longer because of lower velocity but lose their energy higher in the atmosphere. Thus, the heating function maximum moves with the precipitation time from the photosphere ($10 \units{ms}$) to the lower ($17 \units{ms}$) and than upper ($22 \units{ms}$) chromosphere towards the stopping depth of the low energy electrons, where it vanishes.

In a presence of the self-induced electric field (Fig.~\ref{fig:heat-puls10-Bc0}) the heating by collisions becomes smaller than in the purely collisional case because some electrons are reflected by the electric field and do not reach dense plasma and, thus, do not deposit their energy into the ambient plasma. On the other hand, there is a second maximum on the heating function, which is also caused by the energy losses due to Ohmic heating by the self-induced electric field. This maximum does not move but grows in time with more electrons coming to the region with a high electric field (see Fig.~\ref{fig:relax_E}).

As it was shown in Sec.~\ref{sec:var_b} for the magnetic convergence given by Eq.~(\ref{eq:convB3}) only about $10\%$ of the electrons can escape through the loss-cone and heat the deep layers. Thus magnetic convergence substantially reduces the energy deposition at lower atmospheric levels due to the mirroring of the electrons back to the top, which shifts the heating maximum upwards to the corona (Figs.~\ref{fig:heat-puls10-Bc10-noE} and \ref{fig:heat-puls10-Bc10}).

\begin{figure}
  \centering
  \subfloat[collisions]{
    \resizebox{\hsize}{!}{\includegraphics{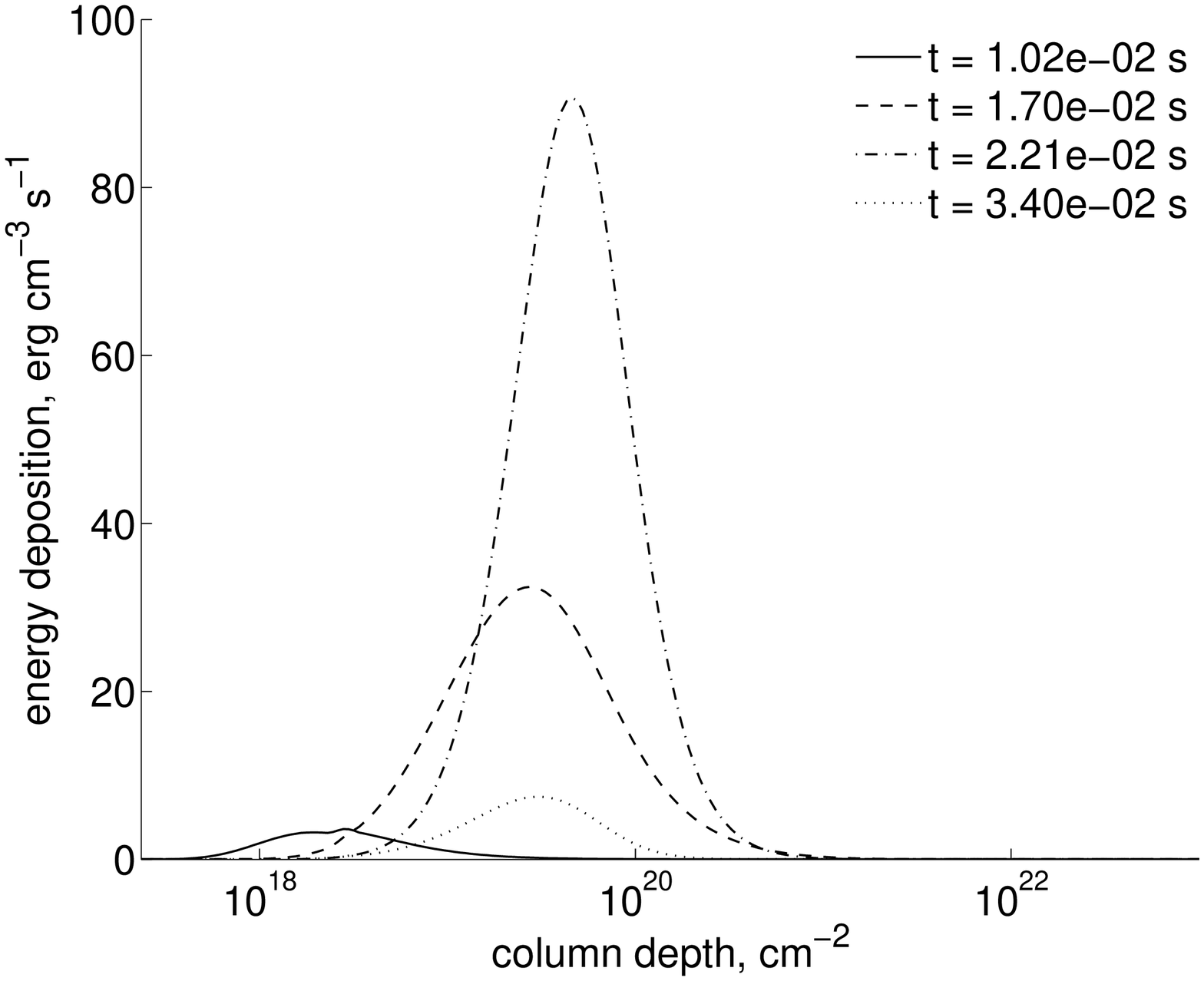}}
    \label{fig:heat-g7-puls10-Bc0-noE}} \quad
  \subfloat[collisions and electric field]{
    \resizebox{\hsize}{!}{\includegraphics{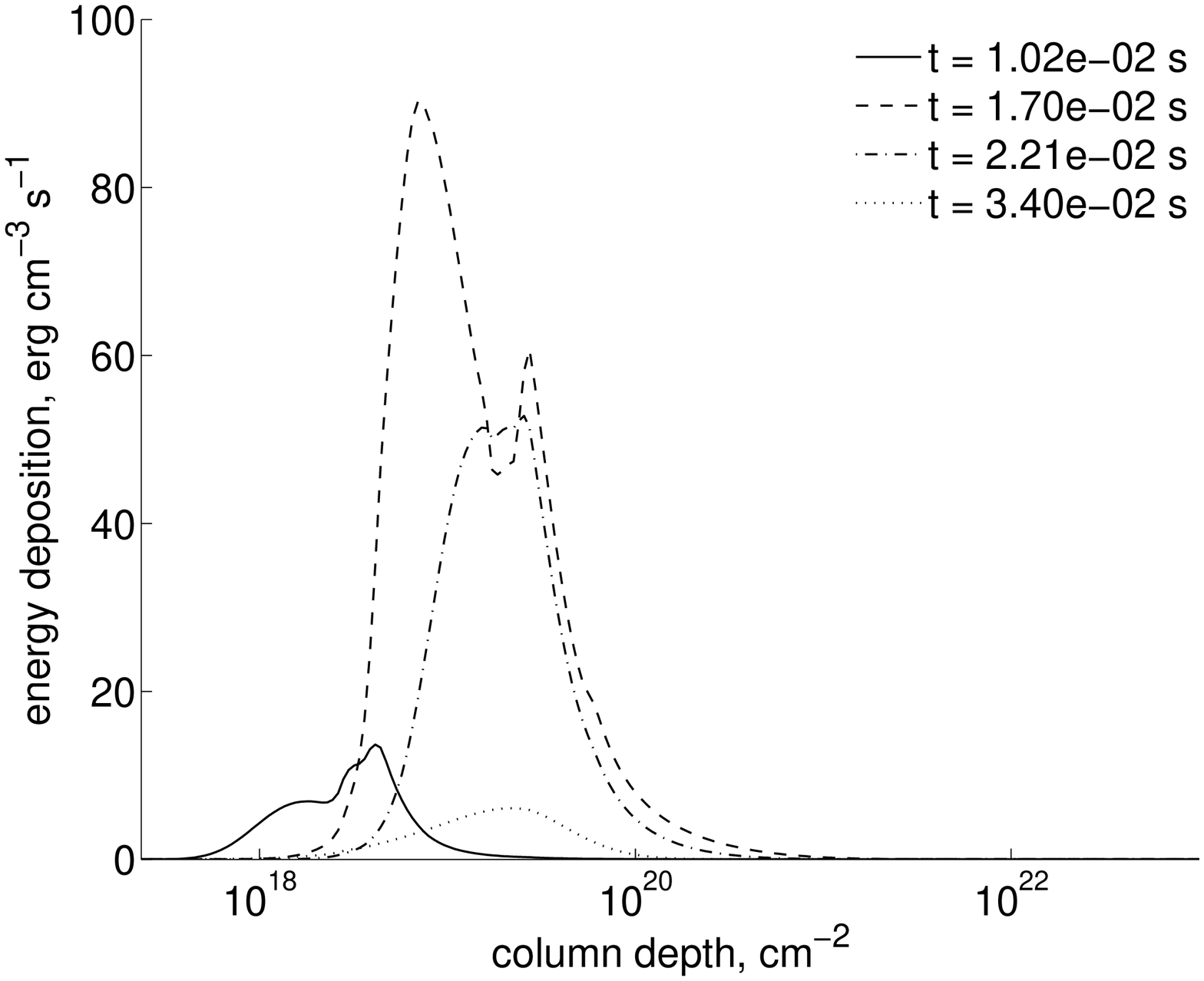}}
    \label{fig:heat-g7-puls10-Bc0}} \\
  \caption{Energy deposition by the impulse of a beam with $\gamma=7$. The other beam parameters are the same as in Fig.~\ref{fig:all-puls10-Bc0}.}
  \label{fig:heat-g7-puls10-all}
\end{figure}

The heating function of a beam with the initial spectral index $\gamma = 7$ is shown in Fig.~\ref{fig:heat-g7-puls10-all}. Contrary to the $\gamma = 3$ beam, the heating peak appears at shallower depths and moves downwards with time. Apparently, this occurs because of the number of high energy electrons being extremely low for a softer beam ($\gamma = 7$), and, thus, the heating that they can produce at greater depths is too low to be noticeable. Also, the heating profile is narrower but higher and its maximum located higher in the atmosphere in comparison with $\gamma = 3$ case. When the electric field is taken into account (Fig.~\ref{fig:heat-g7-puls10-Bc0}) the heating becomes stronger at smaller depths in the corona, in comparison to the pure collisional case (Fig.~\ref{fig:heat-g7-puls10-Bc0-noE}), where it has maximum in the upper chromosphere.

\begin{figure}
  \centering
  \subfloat[collisions]{
    \resizebox{\hsize}{!}{\includegraphics{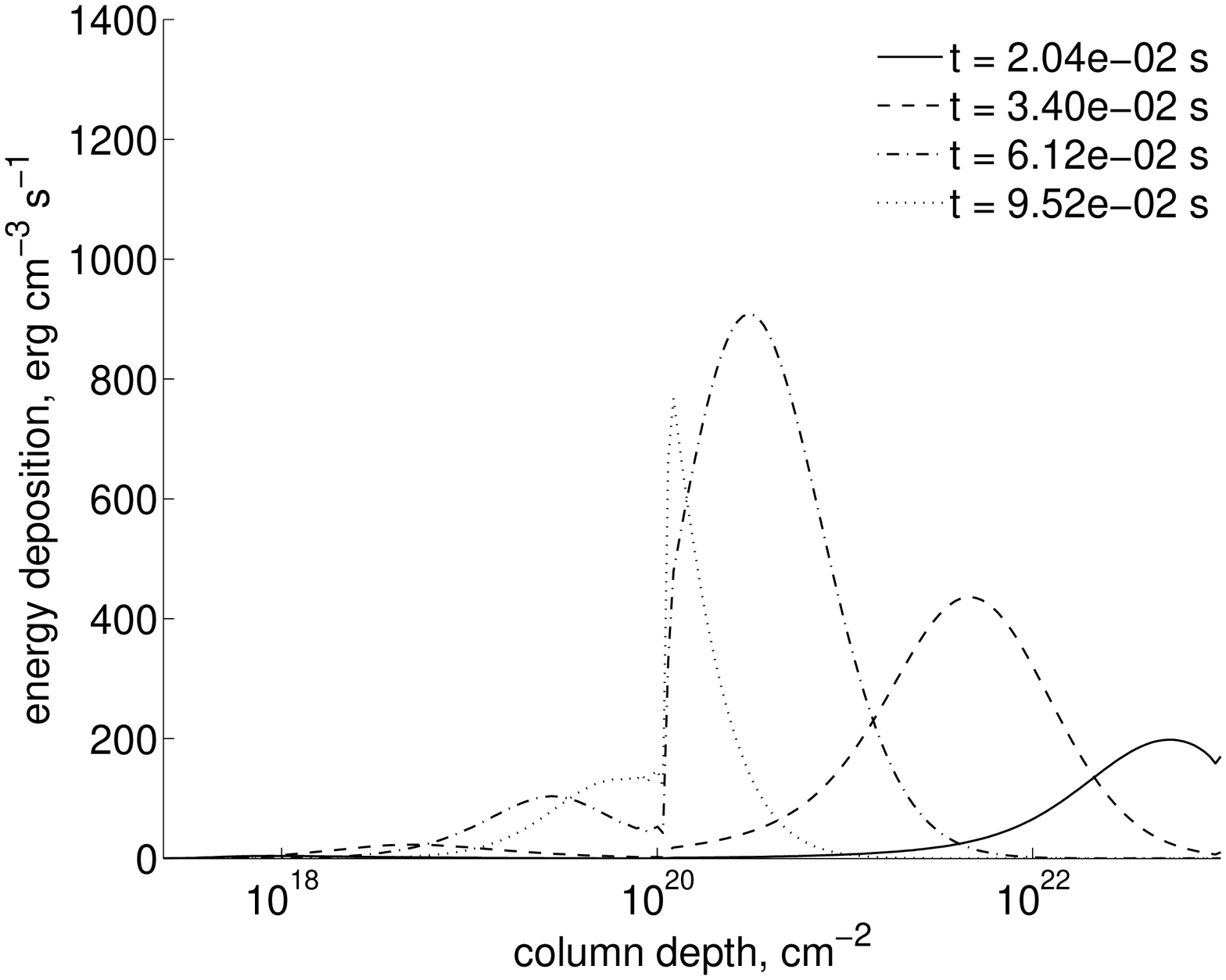}}
    \label{fig:heat-f12-puls10-Bc0-noE}} \quad
  \subfloat[collisions and electric field]{
    \resizebox{\hsize}{!}{\includegraphics{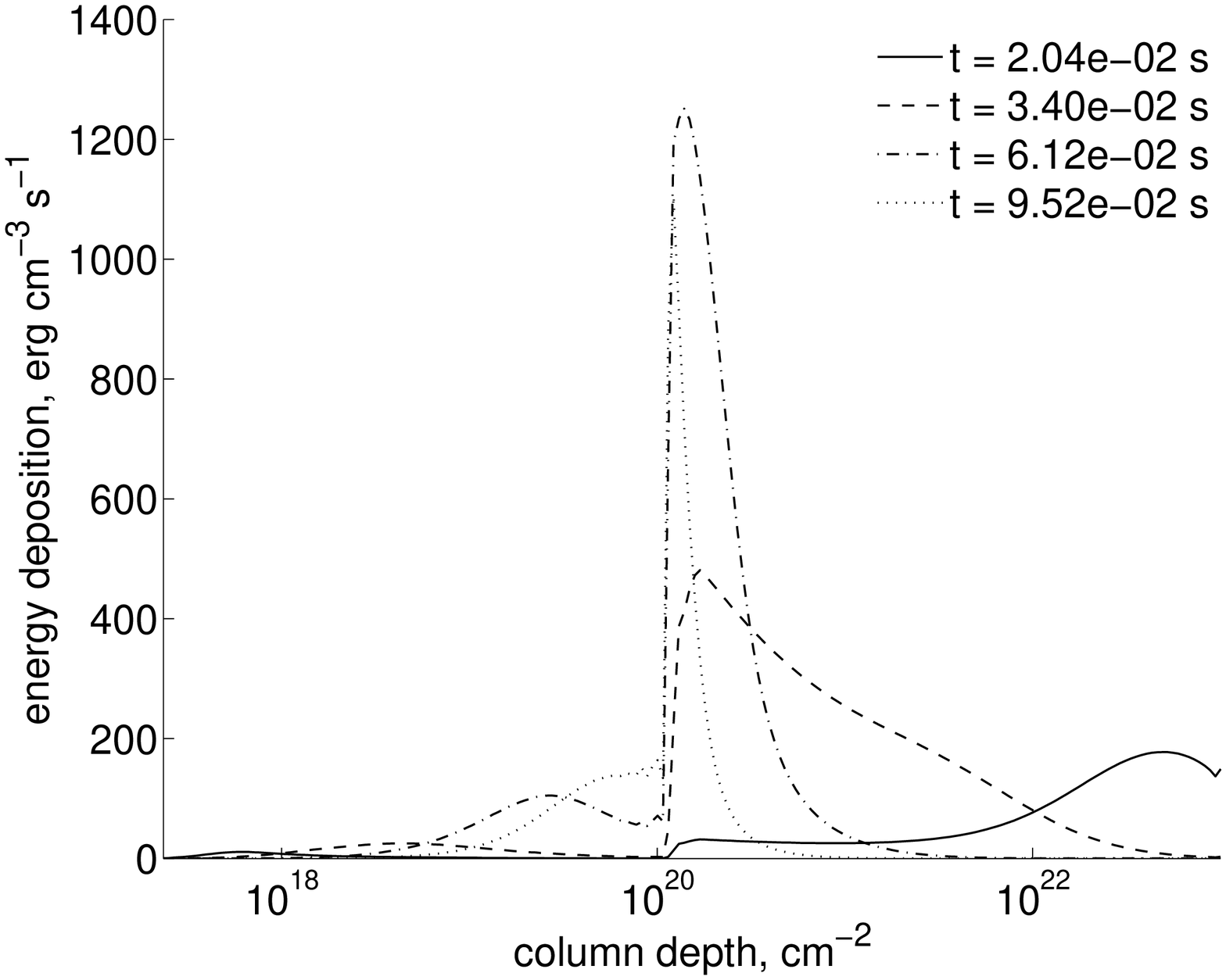}}
    \label{fig:heat-f12-puls10-Bc0}} \\
  \subfloat[collisions and convergence]{
    \resizebox{\hsize}{!}{\includegraphics{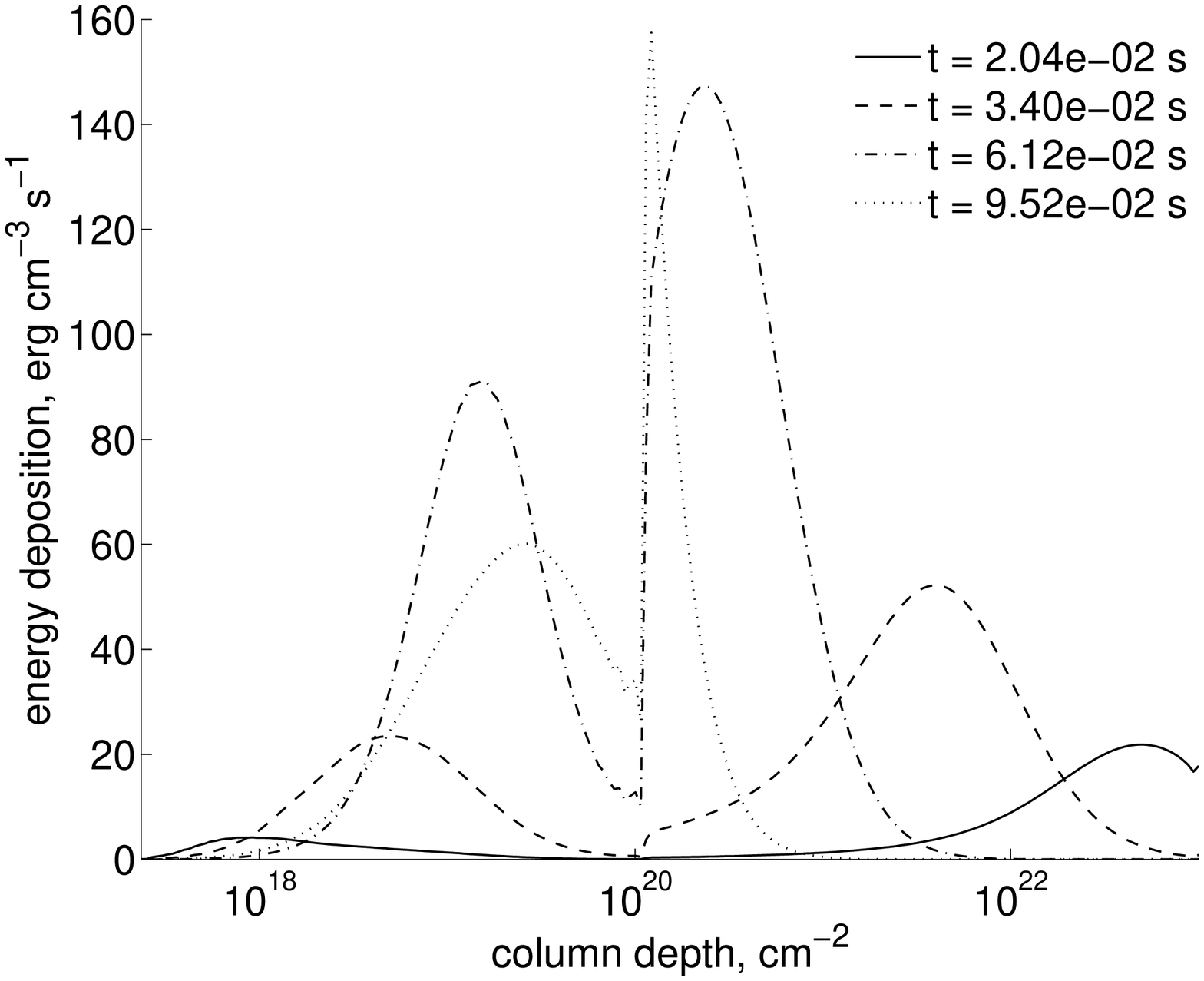}}
    \label{fig:heat-f12-puls10-Bc10-noE}}
  \caption{Energy deposition of the beam impulse with the energy flux $10^{12}\fluxunit$. The other beam parameters are the same as in Fig.~\ref{fig:all-puls10-Bc0} and the magnetic convergence is given by Eq.~(\ref{eq:conv3}).}
  \label{fig:heat-f12-puls10-all}
\end{figure}

A more powerful beam, with energy flux $10^{12}\fluxunit$, obviously deposits more of its energy in the ambient atmosphere (Fig.~\ref{fig:heat-f12-puls10-all}) than a beam with energy flux $10^{10}\fluxunit$. Two maxima are clearly seen on the heating function profile -- one in the chromosphere, another one in the corona. If the magnetic convergence is absent, the chromospheric heating is much stronger (Fig.~\ref{fig:heat-f12-puls10-Bc0-noE}). On the other hand, if the convergence is taken into account, then only about $10\%$ of electrons can reach the chromosphere. Thus, the heating under the transition region is reduced by an order of magnitude, while the coronal heating remains nearly the same as in the case of the constant magnetic field (Fig.~\ref{fig:heat-f12-puls10-Bc10-noE}). Note, that in this case a different hydro-dynamic model is used to estimate the density and temperature of the ambient plasma. The discontinuity at the depth of $10^{20} \percmsq$ (Fig.~\ref{fig:heat-f12-puls10-all}) is caused by a sharp increase of the ambient plasma density (see Fig.~\ref{fig:hd_model}), which apparently corresponds to the transition region.

It can be noted that the evolution time of an electron impulse is longer for a stronger beam. This is a result of a smaller density of the ambient plasma, which leads to a longer relaxation time as was shown in Sec.~\ref{sec:relax}.

\subsection{Bursts of hard X-ray emission}

\begin{figure}
  \centering
  \subfloat[]{
    \resizebox{\hsize}{!}{\includegraphics{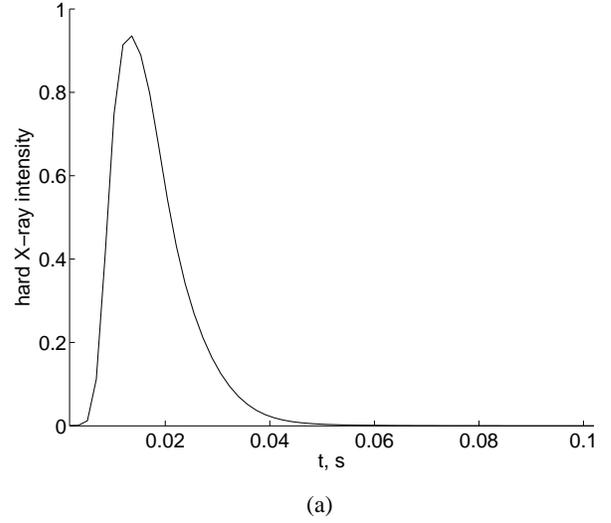}}
    \label{fig:hxrt-puls10-Bc0}} \quad
  \subfloat[]{
    \resizebox{\hsize}{!}{\includegraphics{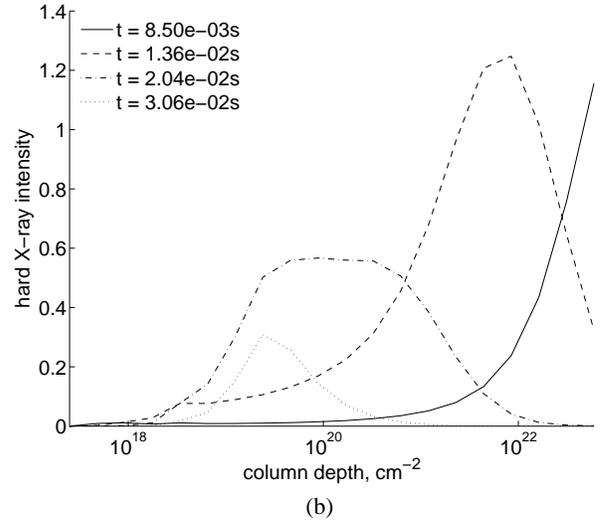}}
    \label{fig:hxrs-puls10-Bc0}} \\
  \caption{Intensity of the hard X-ray emission burst (in arbitrary units): integrated over depth as a function of time (a) and temporal evolution of the spatial profile (b). Beam parameters are the same as in Fig.~\ref{fig:all-puls10-Bc0}.}
  \label{fig:hxr-puls10-Bc0}
\end{figure}

In order to make a comparison with observations we calculate the intensity of hard X-ray emission produced by the injection of a short electron beam. The bremsstrahlung cross-sections are taken in the relativistic form \citep[see][]{Gluckstern53}. Fig.~\ref{fig:hxrt-puls10-Bc0} shows the time profile of hard X-ray intensity produced by the impulse with length $\delta t_\mathrm{e} = 1.7 \units{ms}$. The timescale of the hard X-ray impulse, $\delta t_\mathrm{hxr}$, is about $20 \units{ms}$, which is determined by the relaxation time of the atmosphere $t_\mathrm{r} \approx 70 \units{ms}$ established earlier in Sect.~\ref{sec:relax}. This timescale agrees closely with the observations \citep{Charikov04}. Further simulations show that as long as $\delta t_\mathrm{e} \ll \delta t_\mathrm{hxr}$, the hard X-ray timescale depends only on the atmosphere parameters and does not depend on the length of the initial electron impulse.

Evolution of the spatial profile of the hard X-ray intensity (Fig.~\ref{fig:hxrs-puls10-Bc0}) resembles that of the energy deposition (see Fig.~\ref{fig:heat-puls10-Bc0}). The emission starts at the bottom when high energy electrons reach this depth and gradually moves upwards. After reaching the depth $\sim 2\tento{19}\percmsq$ the intensity the emission decreases and finally the emission vanishes.

\section{Conclusions} \label{sec:concl}

By solving numerically the time-dependent Fokker-Planck equation one is able to study the temporal evolution of the electron beam precipitation in the solar atmosphere and evaluation of the relaxation time required for the beam to reach the stationary regime. For the beam with energy flux $10^{10}\fluxunit$ this relaxation time is $\sim 0.07 \units{s}$ and it becomes longer by a factor of about $3$ (or $\sim 0.2 \units{s}$) for the beam with energy flux $10^{12}\fluxunit$.

The effect of the self-induced electric field during the stationary beam injection is similar to that found in previous studies by \citet{Emslie80, Zharkova06}. In particular, if the electric field is taken into account, then the maximum of the energy deposition profile is shifted upwards making the coronal heating stronger and the chromospheric heating weaker than in the case of pure collisional precipitation.

We considered different models of a converging magnetic field to study the effectiveness of the beam electron refraction by a magnetic mirror. Magnetic field approximations used earlier by \citet{Leach81, McClements92} have the same spatial dependence in the corona and chromosphere. Even if the magnetic field in the photosphere is accepted to be 3 orders of magnitude higher than in the corona, such magnetic profiles are shown to affect only the high energy electrons of the beam. We propose a model where the magnetic field increases exponentially with depth in the corona and becomes constant in the lower chromosphere. Such magnetic field variation can affect the whole energy spectrum of electrons, while the ratio of photospheric/coronal magnetic field is as low as 23. Since the converging magnetic field returns many electrons back to the source, the heating due to collisions and electric field is reduced by $70\%$ in comparison with the constant magnetic field. We also considered the model based on indirect measurements of the magnetic field in the solar atmosphere. Such magnetic field variation can also affect electrons of all energies and reduce the collisional heating by $20\%$ in comparison with the constant magnetic field profile.

The further study is dedicated to the impulsive injections of electrons. In the simulation of impulsive injection the length of the impulse is chosen to be $1.7\tento{-3}\units{s}$, which is much shorter than the relaxation time. It was found that the effect of the electric field is considerably smaller for the short impulse than for the steady injection.

Initial energy spectrum of the injected impulse was power law. It was shown that during the evolution of the impulse the power law index increases in time. For example, if the joint effects of the collisions and magnetic convergence are taken into account, the initial power law index of 3 can increase up to 8 and resembles some quasi-thermal distributions.

The energy deposition profile is shown to depend on the initial power law index. If the energy spectrum is hard ($\gamma = 3$) the heating starts at the bottom end of the system due to the high energy electrons. In the case of soft ($\gamma = 7$) impulse the number of high energy particles is too low to produce any noticeable heating of the deep layers. On the other hand the higher layers are heated more effectively due to the higher number of the low energy electrons in the softer beam.

We also compared the evolution of beams with different intensities. It was found that the difference in this case is mostly cased by the different density and temperature profiles taken from the hydro-dynamic model \citep{Zharkova07} (see Fig.~\ref{fig:hd_model}). For example, the timescale of the impulse evolution is longer the more intense the beam.

If the timescale of the electron impulse is short enough, then the timescale of the hard X-ray emission is determined by the reaction of the atmosphere. Thus, it is of the same order of magnitude as the relaxation time, which, in turn, is of the order of $10 \units{ms}$ and longer. This means that shorter electron impulses can not be detected by the hard X-ray observations.

\begin{acknowledgements}
This research is funded by the Science Technology and Facility Council (STFC) project PP/E001246/1.
\end{acknowledgements}

\bibliographystyle{aa}
\bibliography{/home/taras/Tex/bib_gen}

\end{document}